\documentclass[12pt,a4paper]{article}
\usepackage{graphicx} 
\usepackage{setspace}
\usepackage{amsthm}
\usepackage{amsfonts}
\usepackage{amsmath}
\usepackage{amssymb}
\usepackage[authoryear]{natbib}
\usepackage{doi}
\usepackage{array,booktabs,arydshln,xcolor}
\usepackage{subcaption}
\usepackage[labelsep=quad,skip=3pt]{caption}
\usepackage{graphicx}
\usepackage[page,toc,titletoc,title]{appendix}
\usepackage{siunitx}
\usepackage{authblk}
\onecolumn 

\usepackage{multicol}
\graphicspath{{Figures/}}
\usepackage{pdflscape}

\newtheorem{theorem}{Theorem}%
\newtheorem{lemma}{Lemma}%

\usepackage{dsfont}

\title{Partially Directed Configuration Model with Homophily and Respondent-Driven Sampling}
\author[1]{Alejandro Sepulveda-Pe\~naloza}
\author[2]{Isabelle S. Beaudry}
\affil[1]{Departmento de Epidemiología y Estudios en Salud, Universidad de los Andes, Santiago, Chile}
\affil[2]{Department of Mathematics and Statistics, Mount Holyoke College, Massachusetts, U.S.A}

\begin{document}
\maketitle
\renewcommand{\thefootnote}{\fnsymbol{footnote}}
\renewcommand{\labelitemi}{$\diamond$}

\vspace{1 cc}

\begin{abstract}
  \noindent  Respondent-driven sampling (RDS) is a sampling scheme used in socially connected human populations lacking a sampling frame. One of the first steps to make design-based inferences from RDS data is to estimate the sampling probabilities. A classical approach for such estimation assumes that a first-order Markov chain over a fully connected and undirected network may adequately represent RDS. This convenient model, however, does not reflect that the network may be directed and homophilous. The methods proposed in this work aim to address this issue. The main methodological contributions of this manuscript are twofold: first, we introduce a partially directed and homophilous network configuration model, and second, we develop two mathematical representations of the RDS sampling process over the proposed configuration model. Our simulation study shows that the resulting sampling probabilities are similar to those of RDS, and they improve the prevalence estimation under various realistic scenarios.

\vspace{0.95cc}
\parbox{24cc}{{\it Keywords and Phrases}: Configuration model, Directed network, Homophily, Respondent-driven sampling, Pseudo-sampling proba\-bilities, Hard-to-reach populations.
}
\end{abstract}
\newpage
\section{Introduction}
Since Respondent-Driven sampling (RDS) has been introduced by \cite{Heckathorn97} as a mechanism to sample hard-to-reach human populations, it has been used extensively in a growing number of studies \citep{heckathorn2007, montealegre2013, Berchenko2013ModelingAA, Khan2017OnestepEO, GileBeaudry} and for a multitude of applications such as estimating HIV prevalence in populations of men who have sex with men \citep{Yeka2006, Malekinejad2008, Robineau2020}, in the analysis of migrants populations \citep{Wejnert2010Social, Friberg2014RDSAT}, and fraud prevention \citep{Sosenko2022SmartphonebasedRD}. 

One of the main benefits of RDS is that its implementation does not require a sampling frame since the survey participants are responsible for recruiting additional individuals into the study. Therefore, RDS has been preferred over classical sampling techniques when no such frame exists. However, this apparent advantage is also responsible for an essential drawback of RDS when used for inference purposes: the \textbf{sampling probabilities} are unknown. Consequently, design-based inference from RDS data relies on pseudo-sampling probabilities \citep{Heckathorn97, Salganik2004,gile2011, xinlu2012,fellows2019}.

The main contributions of this work are to propose a novel network model and derive a representation of the RDS process to obtain improved pseudo-sampling probabilities. In particular, our approach considers that RDS is conducted over a partially directed and homophilous network configuration model. RDS is approximated by a sampling scheme with probability proportional to size and without replacement (PPSWOR), also known as successive sampling (SS). The sizes in the SS approximation are selected to reflect the network structure.

Several authors have proposed mathematical representations of RDS to derive the pseudo-sampling probabilities. \cite{Salganik2004} and \cite{volz2008} first modelled RDS as a first-order Markov chain where the states of the stochastic process are the individuals in the network and the transitions, the peer-recruitment. Under many strong assumptions, they show that the stationary distribution of this process is proportional to the participants' out-degrees, that is, the number of people they identify as their contacts in the population. They argue that such stationary distribution may be used as the pseudo-sampling probabilities. However, abundant literature cautions that deviations from the underlying assumptions can lead to significant bias in the estimation \citep{Heckathorn2002, heimer2005, goel2009, Krista_Handcock2010, gile2011, lu2012sensitivity, xinlu2012, beaudry2020}. Therefore, several alternative models have been proposed to relax a subset of the network and sampling assumptions from their original representation.

Among the assumptions often overlooked is the asymmetry of people's relationships. Asymmetry occurs when an individual identifies someone as their contact, but their contact does not reciprocate. Incorporating this network feature in the modelling of the pseudo-sampling probabilities is particularly relevant for RDS because partially directed networks are common in human populations \citep{wallace1966student, south2004friendship, paquette2011use} and ignoring this characteristic may induce bias \citep{lu2012sensitivity}. \citeauthor{xinlu2012} considered nonreciprocating relations while approximating the sampling probabilities. Their methodology adapts the mean-field approach \citep{fortunato2008} to the first-order Markov chain, which leads to the pseudo-sampling probabilities being approximately proportional to the in-degrees, the number of connections toward an individual. 
 
Their work then extends two classical RDS estimators, the \cite{volz2008} and the \cite{Salganik2004} estimators, referred to as the VH and SH estimators, respectively, by replacing the sampling probabilities with the in-degrees. They modified the SH estimator to consider the number of directed connections between individuals with a positive and negative outcome variable, for instance, between infected and uninfected individuals. Their extension of the SH estimator performs especially well when some conditions are met, such as when the sampling fraction is small, and the average in and out degrees are similar for the infected and uninfected individuals.

\cite{xinlu2012} estimators' performance appears to be affected by large sampling fractions, which may be attributable to the use of a with rather than a without replacement RDS sampling representation. In their work, \cite{gile2011} derived pseudo probabilities using a without replacement sampling process. In particular, they model RDS as a self-avoiding random walk over all possible networks of a configuration model with a fixed degree distribution. This approximation is equivalent to sampling with PPSWOR, with unit sizes equal to the out-degrees. This methodology performs well in undirected networks but is not intended for directed networks.

Another widespread feature of the populations studied through RDS is the presence of homophily, that is, the tendency for individuals to form social connections with similar individuals. Ignoring this characteristic can also create bias and increase the variance of the estimators \citep{Crawford2017}. \cite{fellows2019} address this issue with their proposed RDS estimator based on a configuration network model with homophily. They argue that their prevalence estimator is also robust enough to depart from several assumptions, such as seed bias, differential activity, differential recruitment, and short recruitment chains. They show that their estimator results in less bias than some traditional estimators. However, their methodology is also not designed for directed networks.

None of the existing RDS representations simultaneously consider directed networks with homophily and without replacement sampling. Consequently, the two main contributions of this work consist of (1) proposing a novel configuration model for a partially directed network with homophily and (2) proposing two mathematical representations for RDS data based on that network model. 

Our discussion is organized such that Section \ref{notation} presents the notation used throughout this manuscript. The partially directed configuration network model with homophily is introduced in Section \ref{section_2_2}.  
Then, Sections \ref{section_2_3} and \ref{section_2_4} introduce two novel RDS mathematical representations. The first assumes the knowledge of information not traditionally collected in RDS studies, whereas the second relies on estimating that information. We assess the performance of the proposed methods' pseudo sampling probabilities through a simulation study involving a variety of network features in Section \ref{section_simulation_1}. We apply the RDS approximations to a directed network of Wikipedia administrative users' elections in Section \ref{sec:application}. Finally, we summarize our findings in Section \ref{sec:conclusion}.

\section{Basic notation}\label{notation}

We assume that RDS is performed over a network of $N$ individuals. The outcome variable is a binary vector $\boldsymbol{z} \in \{0,1\}^N$. For ease of discussion, we interchangeably use the terms nodes and individuals and refer to $\boldsymbol{z}$ as the infection status for all individuals in the population. We say $z_i=1$ if the individual $i$ is infected, and $z_i=0$ otherwise. The set $\mathcal{Z}^k$ is defined as $\{i: z_i=k \}$, where $k = 0, 1$ and its cardinality is represented by $N^{k}=|\mathcal{Z}^k|$ such that the proportion of infected nodes in the network is $\mu=N^1/N$ (mean or prevalence).

The connections among the nodes are represented by a binary adjacency matrix $Y\in \{0,1\}^{N\times N}$. An entry in such a matrix takes the value one ($y_{ij} = 1$) if nodes $i$ and $j$ are connected and zero otherwise. When $y_{ij} = 1$, we say there is an \textit{edge} between node $i$ and $j$. For an undirected network, $y_{ij} = y_{ji}$ since all edges are reciprocated, whereas for directed networks, $y_{ij}$ may differ from $y_{ji}$.

\section{Attributed configuration model for a partially directed network (ACM)} \label{section_2_2}
This section describes the network model over which RDS is assumed to be conducted. This model aims to capture the partial reciprocity of the relationships and the homophily, that is, the higher probability for similar nodes to connect. A portion of the edges are assumed to be nonreciprocal because we expect some connections in human populations not to be mutual. We propose a configuration network model \citep{BENDER1978, BOLLOBAS_1980,wormald_1980,molloy1995} which incorporates those features. In particular, we propose to extend the partially directed configuration model \citep{2015_britton_conf} to include network homophily on the infection status. 
We refer to this model as the attributed configuration model for a partially directed network (ACM), and we use it to improve the approximation of the sampling probabilities.

\subsection{ACM algorithm} \label{build_configuration_model}
Under a configuration model, networks are generated based on a fixed degree distribution. In the configuration model for a partially directed network \citep{2015_britton_conf}, the authors fixed the distribution of the incoming, outgoing, and undirected stubs (a stub is an unconnected edge). The incoming stubs are then randomly attached to outgoing stubs, and the undirected stubs to other undirected stubs. The authors proved that, under certain conditions, the resulting degree distributions converge to the intended degree distributions. We extend their work by considering a network model for $N$ nodes for which we presume a known infection status and six kinds of stubs: the undirected ($ \cdot \leftrightarrow k$), the incoming ($\cdot \leftarrow k$), and the outgoing ($\cdot \rightarrow k$) stubs from or to nodes in $\mathcal{Z}^k$ where $k = 0,1$. For a node $i$ with known infected status $z_i$, the vector: 
\begin{flalign}
  \boldsymbol{d}^{z_i}_i&=(d^{z_i\leftarrow 1}_i,d^{z_i\leftarrow 0}_i,d^{z_i\rightarrow 1}_i,d^{z_i\rightarrow 0}_i,d^{z_i\leftrightarrow 1}_i,d^{z_i\leftrightarrow 0}_i)  
\end{flalign}
represents the number of stubs of the different types where:
\begin{enumerate}
\item[(a)] $d^{z_i\leftarrow k}_i$: number of incoming stubs from nodes with infection status $k$;
\item[(b)] $d^{z_i\rightarrow k}_i$: number of outgoing stubs to nodes with infection status $k$;
\item[(c)] $d^{z_i\leftrightarrow k}_i$: number of undirected stubs to nodes with infection status $k$,
\end{enumerate}
for $k=0,1$, and $i=1,\ldots,N$. The vector $\boldsymbol{D}^{z_i}_i$ is the random vector associated to $\boldsymbol{d}^{z_i}_i$ and is referred to as the degree of node $i$. The probability that it is equal to $\boldsymbol{d}^{z_i}_i$ given $\boldsymbol{Z}$ is $P(\boldsymbol{D}_i^{z_i}=\boldsymbol{d}_i^{z_i})=p_{\boldsymbol{d}^{z_i}}$. We omit the condition on $\boldsymbol{Z}$ to simplify the notation, that is: 
\begin{align}
p_{\boldsymbol{d}^{z_i}} = P(\boldsymbol{D}_i^{z_i}=\boldsymbol{d}_i^{z_i}|\boldsymbol{Z}=\boldsymbol{z}). 
\end{align}
Unless mentioned otherwise, the distributions in this manuscript are assumed to be conditional on the infection status $\boldsymbol{Z}$. 

The distribution of $\boldsymbol{D}^{z_i}_i$ is written as $F_{\boldsymbol{D}^{z_i}}~\text{for}~i=1, ..., N$. The expected values of the components of $\boldsymbol{D}^{z_i}_i$ are denoted:
\begin{align}
\delta^{z_i\leftarrow k}=&E(D^{z_i\leftarrow k}_i),\\ 
\delta^{z_i\rightarrow k}=&E(D^{z_i\rightarrow k}_i),\\
\delta^{z_i\leftrightarrow k}=&E(D^{z_i\leftrightarrow k}_i).
\end{align}

Assuming there are $N^1$ and $N^0$ infected and uninfected nodes, respectively, the generative process for the ACM is as follows:
\begin{enumerate}
\item[(a)] We create $N$ nodes without stubs and assign $N^1$ of them to a positive infection status and $N^0$ to an uninfected one. Since the order does not matter, without loss of generality, we assign the first $N^1$ nodes to a positive infection status ($z_i=1, i=1,\ldots, N^1$) and the remainder $N^0$ nodes to a negative one ($z_i=0, i=N^1+1,\ldots, N$).
\item[(b)] We independently sample the degree of node $i$ with characteristic $z_i$ from $F_{\boldsymbol{D}^{z_i}}$, for $i = 1, ..., N$.
\item[(c)] We connect pairs of undirected stubs as follows: 

\begin{enumerate}
\item[(i)] Select an undirected stub ($\cdot\leftrightarrow 0$ or $\cdot\leftrightarrow 1$) uniformly at random and identify the characteristic $z_i$ of the node to which it is connected.
\item[(ii)] If the selected stub is from the $z_i \leftrightarrow 0$ type, then pick a stub $z_j \leftrightarrow z_i$ among nodes $j \in \mathcal{Z}^0$. Similarly, if the selected stub is from the  $z_i \leftrightarrow 1$ type, then pick a stub $z_j \leftrightarrow z_i$ among nodes $j \in \mathcal{Z}^1$.
\item[(iii)] Attach the two selected stubs to form an undirected edge.
\item[(iv)] Repeat these steps with the remaining undirected stubs until connecting more undirected stubs is no longer possible. More than one undirected stub may remain disconnected.
\end{enumerate} 

\item[(d)] We connect incoming with outgoing directed stubs as follows: 

\begin{enumerate}
\item[(i)] Select an incoming directed stub ($\cdot\leftarrow 0$ or $\cdot\leftarrow 1$) uniformly at random and identify the characteristic $z_i$ of the node to which it is connected.
\item[(ii)] If the selected stub is from the $ z_i\leftarrow 0$ type, then pick a stub $ z_j \rightarrow z_i$ among nodes $j \in \mathcal{Z}^0$. Similarly, if the selected stub is from the $ z_i\leftarrow 1$ type, then pick a stub $ z_j \rightarrow z_i$ among nodes $j \in \mathcal{Z}^1$.
\item[(iii)] Attach the two selected stubs to form a directed edge.
\item[(iv)] Repeat these steps with the remaining directed stubs until connecting more directed stubs is no longer possible. More than one directed stub may remain disconnected.
\end{enumerate}
\end{enumerate}

After executing the algorithm, we follow \citet{2015_britton_conf}'s methodology to get a simple network. This process includes eliminating unconnected stubs, loops, and parallel edges formed in the algorithm. The following equalities must hold in the simple network:
\begin{align}
s^{0 \leftrightarrow 0} = \sum_{i=1}^{N}d^{z_i\leftrightarrow 0}_i\mathds{1}_{[z_i=0]}&=2 \ell_{00}, \quad \text{for }\ell_{00}\in \mathbb{N} \label{equation_cond_1}\\
s^{1 \leftrightarrow 1}=\sum_{i=1}^{N}d^{z_i\leftrightarrow 1}_i\mathds{1}_{[z_i=1]}&=2 \ell_{11}, \quad \text{for } \ell_{11}\in \mathbb{N}\label{equation_cond_2}
\end{align}

These equations imply that the number of undirected edges in both infection groups must be even. By construction, we also have that:
\begin{align}
s^{0\leftrightarrow 1}=\sum_{i=1}^{N}d^{z_i\leftrightarrow 1}_i\mathds{1}_{[z_i=0]}&=\sum_{i=1}^{N}d^{z_i\leftrightarrow 0}_i\mathds{1}_{[z_i=1]}=s^{1 \leftrightarrow 0}\label{equation_cond_3}\\
s^{0 \rightarrow 1}=\sum_{i=1}^{N}d^{z_i\rightarrow 1}_i\mathds{1}_{[z_i=0]}&=\sum_{i=1}^{N}d^{z_i\leftarrow 0}_i\mathds{1}_{[z_i=1]}=s^{1 \leftarrow 0}\label{equation_cond_4}\\
s^{1 \rightarrow 1}=\sum_{i=1}^{N}d^{z_i\rightarrow 1}_i\mathds{1}_{[z_i=1]}&=\sum_{i=1}^{N}d^{z_i\leftarrow 1}_i\mathds{1}_{[z_i=1]}=s^{1 \leftarrow 1}\label{equation_cond_5}\\
s^{0 \rightarrow 0}=\sum_{i=1}^{N}d^{z_i\rightarrow 0}_i\mathds{1}_{[z_i=0]}&=\sum_{i=1}^{N}d^{z_i\leftarrow 0}_i\mathds{1}_{[z_i=0]}=s^{0 \leftarrow 0}\label{equation_cond_6}\\
s^{1 \rightarrow 0}=\sum_{i=1}^{N}d^{z_i\rightarrow 0}_i\mathds{1}_{[z_i=1]}&=\sum_{i=1}^{N}d^{z_i\leftarrow 1}_i\mathds{1}_{[z_i=0]}=s^{0 \leftarrow 1} \label{equation_cond_7}
\end{align}

We denote as $V^{(N^{z})}_{\boldsymbol{d}^{z}}$ the number of nodes with degree~$\boldsymbol{d}^{z}$ and infection status~$z$ . If we divide $V^{(N^{z})}_{\boldsymbol{d}^{z}}$ by $N^{z}$ we obtain the proportion of nodes that have degree $\boldsymbol{d}^{z}$ among nodes with an infection status of $z$. The empirical degree distribution denoted $F^{(N^{z})}_{\boldsymbol{D}^{z}}$, may be defined using the expected value of those proportions such that:
\begin{flalign}
F^{(N^{z})}_{\boldsymbol{D}^{z}} = \left\{p^{(N^{z})}_{\boldsymbol{d}^{z}}\right\}, \text{ where }~ p^{(N^{z})}_{\boldsymbol{d}^{z}}=E\left(V^{(N^{z})}_{\boldsymbol{d}^{z}}/N^{z}\right).
\end{flalign}

The resulting empirical distribution $F^{(N^{z})}_{\boldsymbol{D}^{z}}$ may differ from the target distribution $F_{\boldsymbol{D}^{z}}$. However, we prove in the supplementary material that $F^{(N^{z})}_{\boldsymbol{D}^{z}}$ converges to $F_{\boldsymbol{D}^{z}}$ 
under specific conditions. 

\subsection{Network configuration model convergence} \label{teorema}
In this section, we present the conditions under which the empirical degree distribution $F^{(N^{z})}_{\boldsymbol{D}^{z}}$ converges to $F_{\boldsymbol{D}^{z}}$.  
The results imply that the proposed ACM converges to the intended degree distribution conditional on the infection status. 

\begin{theorem}\label{teorema_1}
For a fixed $\phi=N^{1}/N^{0}$, we can show that if the mean of the components of $F_{D^1}$ and $F_{D^{0}}$ are finite and related as follows:\\
$ \delta^{1\rightarrow 1}=\delta^{1\leftarrow 1}$, $\delta^{0\rightarrow 0}=\delta^{0\leftarrow 0}$,  $\delta^{0\rightarrow 1}=\phi\delta^{1\leftarrow 0}$,
 $\delta^{0\leftarrow 1}=\phi\delta^{1\rightarrow 0}$,
 $\delta^{0\leftrightarrow 1}=\phi\delta^{1\leftrightarrow 0}$,\\
 then, for $N\to \infty$ and $z=0, 1$:
\begin{multicols}{2}
\begin{enumerate}
\item[(a)] $F^{(N^{z})}_{D^z}\xrightarrow[N\to\infty]{} F_{D^z}$, 
\item[(b)] $V^{(N^{z})}_{\boldsymbol{d}^{z}}/N^{z} \xrightarrow{P} p_{\boldsymbol{d}^{z}}$.
\end{enumerate}
\end{multicols}
\end{theorem}

Theorem \ref{teorema_1} is inspired from \cite{chen2013directed} and \cite{2015_britton_conf} and implies that the empirical distribution converges in probability to $F_{\boldsymbol{D}^{z}}$. However, in the case of the ACM, the distribution is more complex as it has six components rather than three:
\begin{flalign}
    \boldsymbol{D}^{z_i}_i&=(D_i^{z_i\leftarrow 1},D_i^{z_i\leftarrow 0}, D_i^{z_i\rightarrow 1},D_i^{z_i\rightarrow 0}, D_i^{z_i \leftrightarrow 1},D_i^{z_i\leftrightarrow 0}).
\end{flalign}

These components allow the infection status $z$ and the edge directions to be considered. The proof of Theorem \ref{teorema_1} is based on Lemmas \ref{lemma1} to \ref{lemma3}.

\begin{lemma} \label{lemma1}
 $V^{(N^{z})}_{\boldsymbol{d}^{z}}/N^{z} \xrightarrow{P} p_{\boldsymbol{d}^{z}}$ implies $F^{(N^{z})}_{\boldsymbol{D}^{z}}\to F_{\boldsymbol{D}^{z}}$ as $N^z\to \infty$.
\end{lemma}

\begin{lemma} \label{lemma2}
Let $M^{(N^{z_r})}_r$ be the indicator variable equal to 1 if the node $r$ with an infection status $z_r$ has had its degree modified in the process of creating a simple configuration model network, and zero otherwise. For any arbitrary $r$ such that $\boldsymbol{d}^{z_r}$ and $N^{z_r}\to \infty$: 
$$\text{If } P(M^{(N^{z_r})}_r=0|\boldsymbol{D}_r^{z_r}=\boldsymbol{d}_r^{z_r})\to 1 \text{ then } \frac{V^{(N^{z_r})}_{\boldsymbol{d}^{z_r}}}{N^{z_r}}\xrightarrow{P} p_{\boldsymbol{d}^{z_r}} $$
\end{lemma}

\begin{lemma} \label{lemma3}
Let $\{X_m\}$ be a sequence of non-negative random variables, and let $X$ be a non-negative random variable. Also, let $0<a<\infty$ be a real number. If $X_m \xrightarrow{D} X$ as $m\to \infty$, $\lim_{m\to\infty} E[X_m]\leq a$ and $E[X] = a$, then:
$$lim_{m\to \infty} E[X_m] = a.$$
\end{lemma}

 \noindent The proofs of Theorem \ref{teorema_1} and Lemmas 1 and 2 are presented in the supplementary material. Lemma~\ref{lemma3} is proven by \cite{2015_britton_conf}.

Our proposed RDS representation developed in Section \ref{section_2_3} assumes the sampling process occurs over the ACM. In particular, the representation is derived from a random walk over the space of all ACM with fixed degree distribution and infection status.

\section{Successive Sampling proportional to partial in-degrees} \label{section_2_3}
 This section describes the successive sampling (SS) approximation to RDS over an ACM. Under SS, units in the population are sampled sequentially, without replacement, and with probability proportional to a given unit size. \cite{gile2011} first introduced SS as a representation of RDS to reduce the estimation bias with large sampling fractions. The author showed that the participants' out-degrees may be used as the SS unit sizes when the sampling is performed over undirected networks. \cite{xinlu2012} used the in-degrees as the SS unit sizes for directed networks. Although these unit sizes are a suitable approximation for directed networks (see proof in the supplementary material), they do not consider the network homophily. Our proposed SS unit sizes consider homophily and partially directed networks.

We propose to model RDS as a self-avoiding random walk over the space of all networks of a given ACM. The transitions represent the act of someone recruiting another individual. Therefore, a transition between $i$ and $j$ may occur if there are any outgoing or undirected stubs from $i$ and any incoming or undirected stubs to $j$. To ensure the recruitment accurately reflects the proportion of edges between and within infection groups, the transition probability to any $j \ne i$ is proportional to its number of incoming and undirected stubs from $z_i$ ($\cdot \leftarrow z_{i}$ and $\cdot \leftrightarrow z_{i}$). 

\begin{figure}[th]
\centering
\includegraphics[width=5in]{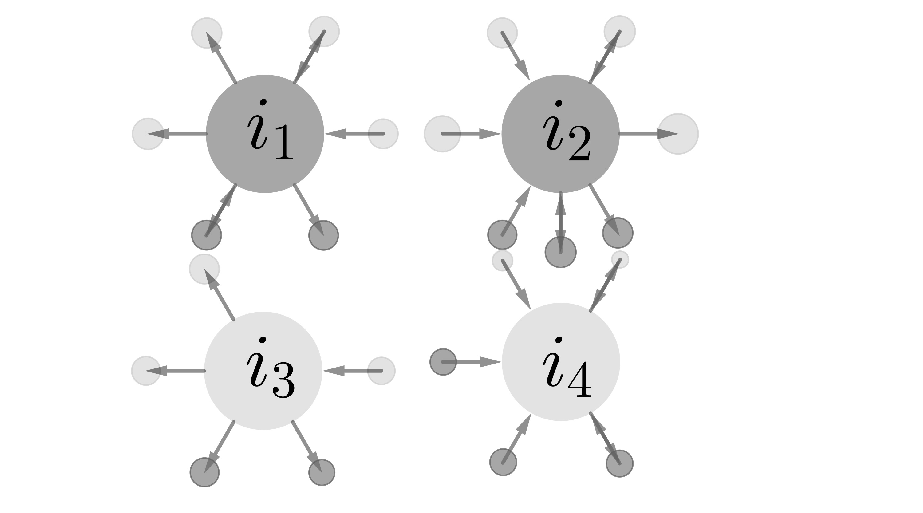}
\caption{Four nodes with index $i_1, i_2, i_3, i_4$ with their associated stubs. Nodes $i_1$ and $i_2$ are infected (dark grey), while nodes $i_3$ and $i_4$ are uninfected (light grey).}
\label{fig:toysamp}
\end{figure}

For instance, let us assume that node $i_1$ from Figure~\ref{fig:toysamp} has to recruit one node. Let us further assume that the dark grey circles in  Figure~\ref{fig:toysamp} indicate a positive infection status and the light grey circles a negative one. Since $i_1$ is infected, it may only recruit nodes with incoming ($\cdot \leftarrow 1 $) or undirected ($ \cdot \leftrightarrow 1$) stubs from infected nodes. Therefore, $i_1$ may recruit $i_2$ since $d^{1\leftarrow 1}_{i_2} + d^{1\leftrightarrow 1}_{i_2} = 1+ 1= 2 > 0$
or $i_4$ since $d^{0\leftarrow 1}_{i_4} + d^{0\leftrightarrow 1}_{i_4} = 2+ 1= 3 > 0$. However, $i_1$ may not recruit $i_3$ since $d^{0\leftarrow 1}_{i_3} + d^{0\leftrightarrow 1}_{i_3} = 0$. In this example, the probability of transitioning from $i_1$ to $i_2$ and $i_4$ is 2/5 and 3/5, respectively. That is, the probability that $i_1$ select $i_2$ and $i_4$ is proportional to their number of incoming and undirected stubs from an infected node.

More formally, if $G_n$ denotes the state of the random walk at step $n$, the transition probability is given by:
\begin{align}
P\left(G_j=g_j \right. & \left. \mid G_1,G_2, \ldots, G_{j-1}=g_1, g_2, \ldots, g_{j-1}\right) \nonumber\\
& = \begin{cases}
\displaystyle\frac{d^{in,z_{g_{j-1}}}_{g_j}}{\sum_{k=1}^{N}d^{in,z_{g_{j-1}}}_{k}-\sum_{k=1}^{j-1}d^{in,z_{g_{j-1}}}_{k}}, & g_j \notin g_1, \ldots, g_{j-1} \\
0, & g_j \in g_1, \ldots, g_{j-1}
\end{cases} \label{transition_prob}
\end{align}
where $d^{in,k}_i = d^{z_i\leftarrow k}_i + d^{z_i\leftrightarrow k}_i$ for $i = 1, ..., N$ and $k=0,1$ are referred to as the \textbf{partial in-degrees}.

We call this approximation the SS proportional to partial in-degrees ($SS_{pi}$), and we observe that every transition depends on the infection status of the recruiting node and the partial in-degrees of all nodes in the network. The simulation study and application discussed in Section \ref{section_simulation_1} show that the sampling probabilities obtained from this process are a good approximation to RDS sampling probabilities under the scenarios we considered. 

\section{Successive Sampling proportional to approximate partial in-degrees} \label{section_2_4}
Section \ref{section_2_3} proposes an SS representation of RDS based on the partial in-degrees knowledge. However, this information is rarely available in traditional RDS applications. In this section, we approximate the partial in-degrees through the total number of edges between and within infection groups.

We first define the total number of edges between and within infection groups as follows: 
\begin{align}
E_{11}&=\sum_{i,j}y_{ij}z_{i}z_{j}, \\ 
E_{10}&=\sum_{i,j}y_{ij}z_{i}(1-z_{j}), \\ 
E_{01}&=\sum_{i,j}y_{ij}(1-z_{i})z_{j}, \\ 
E_{00}&=\sum_{i,j}y_{ij}(1-z_{i})(1-z_{j}). \label{bloques} 
\end{align}

These quantities are the number of edges $y_{ij}$ connecting people of specific infection status. For instance, $E_{11}$ represents the connections among infected people, whereas $E_{10}$ is the number of edges from infected to uninfected. Assuming $E_{11}, E_{10}, E_{01}$ and $E_{00}$ are known or could be reasonably estimated, the proportion of incoming edges toward the infection status $l$ ($E_{0,l}+E_{1,l}$) that originated from the infection status $k$ ($E_{k,l}$) is:
\begin{equation} \label{cross_in}
R^{in}_{l, k}=\frac{E_{k,l}}{E_{0,l}+E_{1,l}}.
\end{equation}

Under the ACM, the expected partial in-degree for a node $j \in \mathcal{Z}^{z_j}$ may be approximated using the $R^{in}_{l, k}$ ratios. In particular, we may approximate them as a proportion of the in-degree, such that:
\begin{align}
E[d^{in,1}_j] & \approx R^{in}_{z_j, 1}\cdot d^{in}_j,\\
E[d^{in,0}_j] & \approx R^{in}_{z_j, 0}\cdot d^{in}_j.
\end{align}

To obtain an approximation of the transition probabilities, we propose to replace the partial in-degrees in Equation (\ref{transition_prob}) by their approximated expected values, such that:
\begin{align}
P&\left(G_j=g_j \mid G_1,G_2, \ldots, G_{j-1}=g_1, g_2, \ldots, g_{j-1}\right) \nonumber\\
&\approx \begin{cases}
\displaystyle\frac{R^{in}_{z_{g_{j}} z_{g_{j-1}}}d^{in}_{g_j}}{\sum_{k=1}^{N}R^{in}_{z_k z_{g_{j-1}}}d^{in}_{k}-\sum_{k=1}^{j-1}R^{in}_{z_{g_k} z_{g_{j-1}}}d^{in}_{g_k}}, & g_j \notin g_1, \ldots, g_{j-1} \\
0, & g_j \in g_1, \ldots, g_{j-1}.\end{cases}
\end{align}

We call this process the SS proportional to the approximated partial in-degrees~($SS_{pa}$).

\section{Simulation study}\label{section_simulation_1}

We have proposed two mathematical representations of RDS to obtain pseudo-sampling probabilities for samples collected over partially directed networks with homophily. This section compares the performance of the pseudo-inclusion probabilities from these two RDS models, the SS with partial in-degree ($SS_{pi}$) and the SS with approximated partial in-degree ($SS_{pa}$), with the performance of the SS with in-degree ($SS_{in}$) and the random walk with-replacement with probability proportional to the in-degree (WRPI). Those four approximations are benchmarked against the simulated RDS inclusion probabilities under various network and sampling conditions.

This section describes the design and results from the simulations study. In particular, we discuss how the networks were generated (Section~\ref{network_simulation}). Then, we explain how the samples were simulated and the pseudo-inclusion probabilities calculated (Section~\ref{samling_procedures}). Finally, we assess the performance of the inclusion probabilities and the prevalence estimators in Sections~\ref{comparison_PI} and~\ref{simresults_prevalence}, respectively.

\subsection{Network simulation}\label{network_simulation}

This section explains how we simulated the networks from which we extracted samples. The principal characteristics we aimed to capture are the network homophily~($h$), the attractiveness ratio~($m$), and the activity ratio~($w$).

Homophily is the tendency for nodes to connect with others with similar characteristics. We define homophily with respect to the infection status as follows:
\begin{equation}\label{h2}
h=\frac{P(Y_{ij}=1|Z_i=Z_j=1)}{P(Y_{ij}=1|Z_i\neq Z_j)}.
\end{equation}
As shown in table \ref{tab:01}, the expected strength of the simulated homophily is either one or five, corresponding to a random attachment and high homophily level scenarios, respectively \citep{Krista_Handcock2010, beaudry2020}.
\begin{table}[!t]
\caption{Parameters used to simulate the networks: homophily ($h$), attractiveness ratio ($m$), activity ratio ($w$), and proportion of directed edges ($\alpha$).  \label{tab:01}}
\begin{tabular*}{\columnwidth}{@{\extracolsep\fill}cccc@{\extracolsep\fill}}
\hline
$m$ & $w$ & $h$ & $\alpha$\\ 
\hline
0.8 & 0.8 & 1 & 0.8\\
1 & 1& 5& 0.2\\
2 & 2& & \\\hline
\end{tabular*}
\end{table}

The simulation study also assesses the sensitivity of the results to various levels of attractiveness ratio ($m$) and activity ratio ($w$). The attractiveness ratio compares the average in-degree of infected individuals to that of uninfected individuals, while the activity ratio compares their average out-degree \citep{gile2011, xinlu2012}:
\begin{flalign}
m = \bar{d}^{in}_1 / \bar{d}^{in}_0, \text{ and } w = \bar{d}^{out}_1 / \bar{d}^{out}_0,\label{mw}
\end{flalign}

where $\bar{d}^{in}_z$ and $\bar{d}^{out}_z$ are the averages in and out-degree, respectively, for nodes with characteristic $z=0,1$. The parameters $m$ and $w$ are set to 0.8, 1, or 2 (${m, w = 0.8, 1, 2}$). Also, the overall average degree equals 10 ($\lambda = 10$), and the proportion of directed edges equals 0.2 or 0.8 ($\alpha = 0.2, 0.8$).

We generated 36 networks for all combinations of these parameters.
The networks contained $N=1500$ nodes, of which $N^1=300$ were infected, and ${N^0=1200}$ were not; that is, we set the proportion of infected nodes to ${\mu=N^1/N=20\%}$.

To create the network adjacency matrix $Y$, we drew each edge independently from a Bernoulli distribution with probability conditional on the infection status of the nodes included in the dyad, such that:
\begin{align}\label{pkl}
p_{kl}=P(Y_{ij}=1|Z_{i}=k, Z_{j}=l), \text{ for     } k, l = 0,1.
\end{align}

The probabilities were determined to be consistent with the network's parameters. In particular, the probabilities are equal to:
\begin{align}\label{prob_enlaces}
  \boldsymbol{p}_{00}&=P(Y_{ij}=1|Z_{i}=Z_{j}=0) = \frac{E_{00}^s}{N^{0}(N^{0}-1)}, \\
  \boldsymbol{p}_{11}&=P(Y_{ij}=1|Z_{i}=Z_{j}=1) = \frac{E_{11}^s}{N^{1}(N^{1}-1)}, \hskip 1cm \\
  \boldsymbol{p}_{10}&=P(Y_{ij}=1|Z_{i}=1,Z_{j}=0) = \frac{E_{10}^s}{N^{1}\cdot N^{0}}, \hskip 1.3cm \\
  \boldsymbol{p}_{01}&=P(Y_{ij}=1|Z_{i}=0,Z_{j}=1) = \frac{E_{01}^s}{N^{1}\cdot N^{0}}.
\end{align}

where $E_{kl}^s, k,l = 0,1$, are the total number of edges from $i\in Z^k$, to $j\in Z^l$ in the simulations, which are determined such that:
\begin{align}
E_{00}^s  =& \left(\frac{1}{1+\phi\cdot m}+\frac{1}{1+\phi\cdot w}+\frac{1}{1+1/H}-1\right)\left(\frac{\lambda \cdot N(H+1)}{2H+1}\right), \hskip 1cm \label{E00s}\\
E_{11}^s  =& \frac{\lambda\cdot N-E_{00}^s}{1+1/H},\label{E11s}\\
E_{10}^s  =& \frac{\lambda\cdot N}{1+\phi\cdot m}-E_{00}^s, \hskip 7.3cm \label{E10s}\\
E_{01}^s  =&\frac{\lambda\cdot N}{1+\phi\cdot w}-E_{00}^s,\label{E01s}
\end{align}

where $H=h\cdot (N^{1}-1)/(2\cdot N^{0})$, and $\phi=N^{1}/N^{0}$. Finally, we randomly selected $100(1-\alpha)$\% edges in each infection block to form undirected edges. The algorithm to distribute the edges in each block is described in the supplementary material.

\subsection{Sampling procedures and approximated inclusion  probabilities}\label{samling_procedures}

Once we generated the networks under the thirty-six scenarios using the parameters shown in Table \ref{tab:01}, we drew samples according to the sampling methods discussed in this manuscript: RDS, WRPI, $SS_{in}$, $SS_{pi}$, and $SS_{pa}$. The main objective is to compare the inclusion probabilities of the four RDS representations for directed graphs with those of RDS.

For each network, we selected 500 samples of size 200, 500, 750, and 1125 through WRPI, $SS_{in}$, $SS_{pi}$, and $SS_{pa}$, and 1000 RDS samples of the same sizes. The different sample sizes are meant to analyze the effect of the sampling fraction on the performance of the approximations. In particular, the successive sampling approach (SS) is expected to perform better than a with-replacement representation for large sampling fractions \citep{gile2011}. We verify this claim in our simulations.

To initiate the RDS samples, we selected ten seeds at random. From each node, we randomly chose two of their contacts. The process stopped when the sample size was reached.

We calculated the relative frequency of each node across the samples of a given scenario to approximate the nodes' inclusion probability for each sampling procedure: 
\begin{equation}\label{eq:pihat}
\hat{\pi}_i^{app} = \frac{1}{n_{samp}} \sum_{j = 1}^{n_{samp}} \mathds{1}_{[i \in \mathcal{S}_j^{app}]},
\end{equation}
where $\mathds{1}_{[i \in \mathcal{S}_j^{app}]}$ equals one if $i$ is in $\mathcal{S}_j^{app}$, the set of all nodes included in the $j$-th sample generated with the approximation $app \in \{WRPI, SS_{in}, SS_{pi}, SS_{pa}\}$, or zero otherwise, and where $n_{samp} = 500$ is the number of samples. The inclusion probabilities are compared to RDS' simulated probabilities, calculated similarly, in Section \ref{comparison_PI}.

\subsection{Comparison of the inclusion probabilities} \label{comparison_PI} 
To assess the accuracy of the RDS approximations, we calculated the mean absolute relative error (MARE) for each network scenario and sampling method, where the MARE is defined such that:
\begin{equation}\label{MARE}
    MARE=\frac{1}{N}\sum_{i=1}^N\frac{|\hat{\pi}^{app}_i-\hat{\pi}^{rds}_{i}|}{\hat{\pi}^{rds}_i}.
\end{equation} 
This quantity measures the average relative discrepancy between the approximated and RDS simulated inclusion probabilities. 

\begin{figure}[!ht]
    \centering
    \includegraphics[width = 11cm]{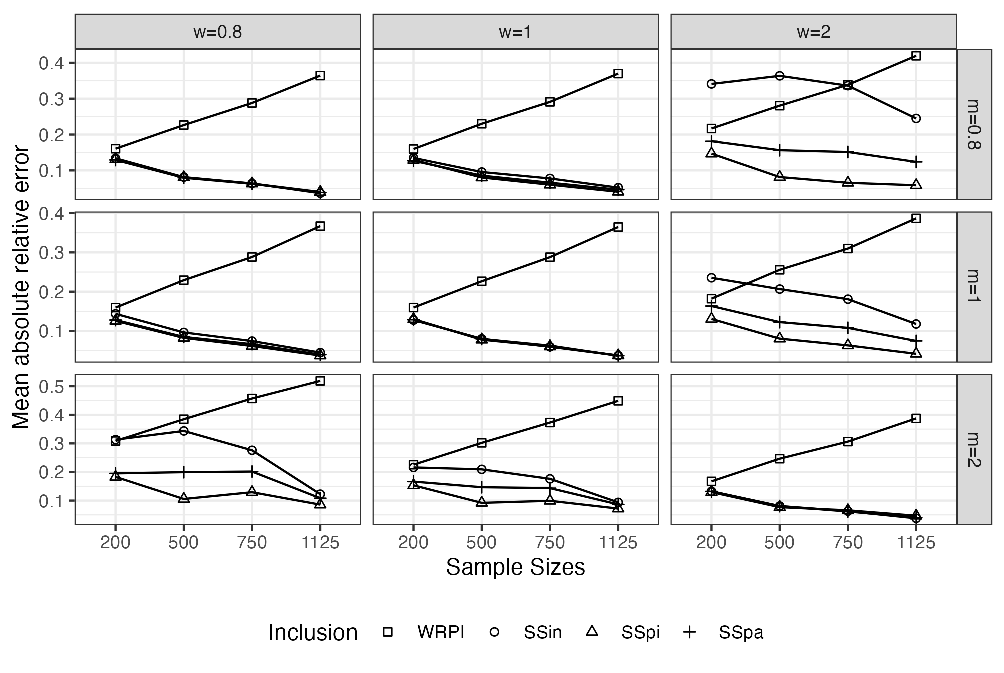}
    \caption{Mean absolute relative error for nine scenarios where the homophily is high ($h=5$), and the proportion of directed edges is 0.2.}
    \label{fig:relative_error_1}
\end{figure}

Figure \ref{fig:relative_error_1} shows the results for the sample sizes $200, 500, 750,$ and 1125 for networks with a high homophily ($h = 5$) and with a low proportion of directed edges ($\alpha = 0.2$). The plot is organized into nine regions, each representing a different network scenario varying the level of activity ratio ($w$), which increases from left to right, and attractiveness ratio ($m$), which increases from top to bottom. The results with random attachment ($h=1$) are presented in the supplementary material. 

We observe in Figure \ref{fig:relative_error_1} that the MARE for our RDS approximations, $SS_{pi}$ and $SS_{pa}$, are either lower or similar to the alternative approaches $SS_{in}$ and WRPI. In particular, the differences between $SS_{in}$ and $SS_{pi}$ and $SS_{pa}$ are especially significant when $m \ne w$ and for small sample sizes. This behaviour may be attributed to the fact that both $SS_{pi}$ and $SS_{pa}$ consider the infection status in the recruitment process, and when $m\ne w$, one infection group tends to be disproportionately selected in the samples. The gap, however, decreases as the sample size increases since the three approximations are based on without-replacement sampling and, therefore, are less subject to the finite population bias. As such, the inclusion probabilities under these three methods improve as we collect additional information. The finite population effect may explain the difference with WRPI. When the sampling fraction is small ($n = 200$), the WRPI is closer to the SS-based approximations. However, as the sampling fraction increases, the WRPI inclusion probabilities worsen because the WRPI models RDS as a with-replacement sampling process.

When comparing our two proposals, we observe that $SS_{pi}$ yields a lower or equivalent error rate to $SS_{pa}$ in all the scenarios considered. The reduced errors occur since $SS_{pi}$ relies on the actual values of partial degrees, whereas $SS_{pa}$ relies on approximating these quantities. This approximation through the edge count proportions $R_{kl}^{in}, k, l = 0,1$ does not yield the same accuracy when $m \ne w$. However, the loss in accuracy tends to be relatively small in most cases, and $SS_{pa}$ offers the advantage that it is based on information that may be easier to obtain or estimate.

\subsection{Prevalence comparison}\label{simresults_prevalence}
The main objective of getting better pseudo-inclusion probabilities with RDS data is to improve the prevalence estimation of a binary characteristic, such as the infection status. Therefore, we used the pseudo-inclusion probabilities from the simulation study in the Hájek estimator to estimate the prevalence, such that:
\begin{equation}
\hat{\mu}^{app}=\frac{\displaystyle\sum_{i=1}^{n}\frac{z_{i}}{\hat{\pi}_i^{app}}}{\displaystyle\sum_{i=1}^{n}\frac{1}{\hat{\pi}_i^{app}}},
\end{equation} 
where $app \in \{WRPI, SS_{in}, SS_{pi}, SS_{pa}\}$. For comparison purposes, we also included the estimator proposed by \cite{xinlu2012}, $SH_{in}$, which also relies on the WRPI pseudo-inclusion probability. 
We calculated each scenario's Root Mean Square Error (RMSE) for each estimator.

\begin{figure}[!ht]
\centering
\includegraphics[width = 11cm]{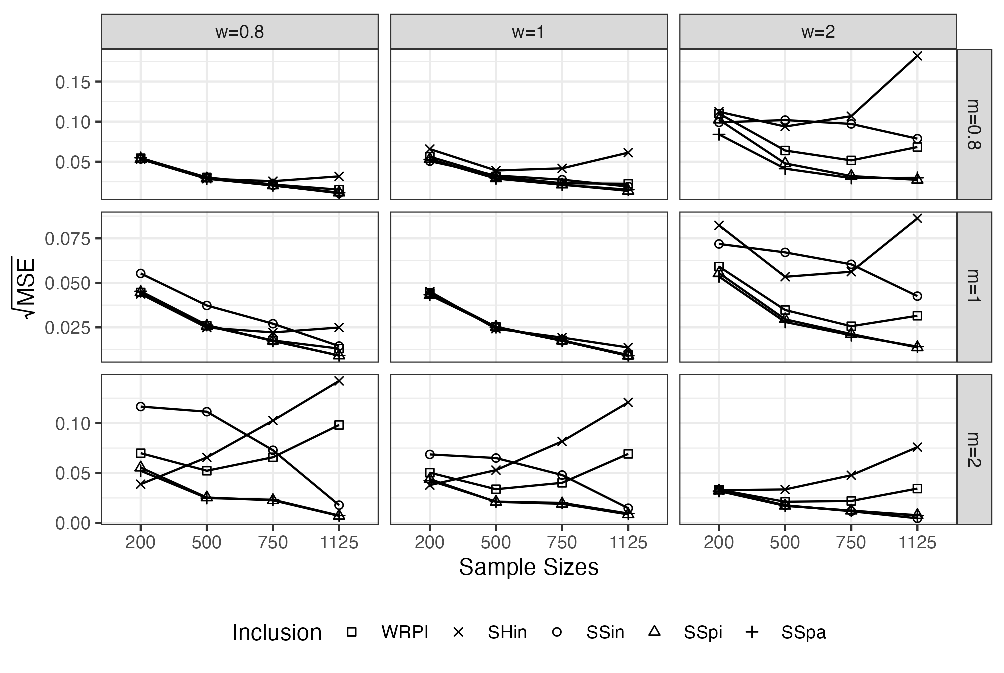}
\caption{The root mean square error for the prevalence estimators at various sample sizes calculated with the simulated inclusion probabilities based on the approximation WRPI, $SS_{in}$, $SS_{pi}$ and $SS_{pa}$. The figure also includes the $SH_{in}$ estimator. The proportion of directed edges is 0.2, and the homophily is equal to 5.}
\label{MSE_prev_h5}
\end{figure}

Figure \ref{MSE_prev_h5} shows the RMSE for the different Hájek estimators with inclusion probabilities given by WRPI, $SS_{in}$, $SS_{pi}$ and $SS_{pa}$, and the $SH_{in}$ estimator at various sample sizes (200, 500, 750, and 1125). Similarly to Figure~\ref{fig:relative_error_1}, Figure~\ref{MSE_prev_h5} also presents nine scenarios for the different combinations of $m$ and $w$. The results are based on a proportion of directed edges equal to 0.2 and a homophily level of five. The results with a high proportion of directed edges, $\alpha=0.8$, are not shown in this manuscript since they are similar to those with $\alpha=0.2$. Still, those with low homophily ($h=1$) are presented in the supplementary material.

From Figure \ref{MSE_prev_h5}, we observe that the results are consistent with the inclusion probabilities results. The Hájek estimators based on $SS_{pi}$ and $SS_{pa}$ have similar or lower RMSE than the alternative estimators in almost all scenarios. Estimators based on $SS_{pi}$ and $SS_{pa}$ provide a larger competitive advantage over WRPI and $SH_{in}$ in highly homophilous networks when $m \ne w$ and for large sampling fractions. 

We note that $SH_{in}$ is among the estimators with the lowest RMSE in many scenarios when $n = 200$.
However, as the sampling fraction increases, its performance deteriorates since the finite population effect affects the WRPI approximation used in some of the components of the $SH_{in}$ estimator. 

Also, similarly to what we observed in the comparison of the inclusion probabilities, the estimators based on $SS_{pi}$ and $SS_{pa}$ display better statistical properties than that based on the $SS_{in}$ when $m \ne w$ and for small sampling fractions.

Under some scenarios, such as when $w =2, m = 0.8$ and $n$ is at most 750, the estimator based on $SS_{pa}$ yields a lower RMSE than the estimator based on $SS_{pi}$ despite the inclusion probabilities based on $SS_{pi}$ being closer to the RDS inclusion probabilities on average as discussed in Section \ref{comparison_PI}. This situation arises since the error rates on the inclusion probability are nonuniform across the nodes. Also, the prevalence estimation is more heavily affected by errors on small inclusion probabilities than large ones. 
 
Therefore, the larger error rates on small inclusion probabilities under $SS_{pi}$ led to this counterintuitive result.

\begin{figure}[!ht]
\centering
\includegraphics[width = 11cm]{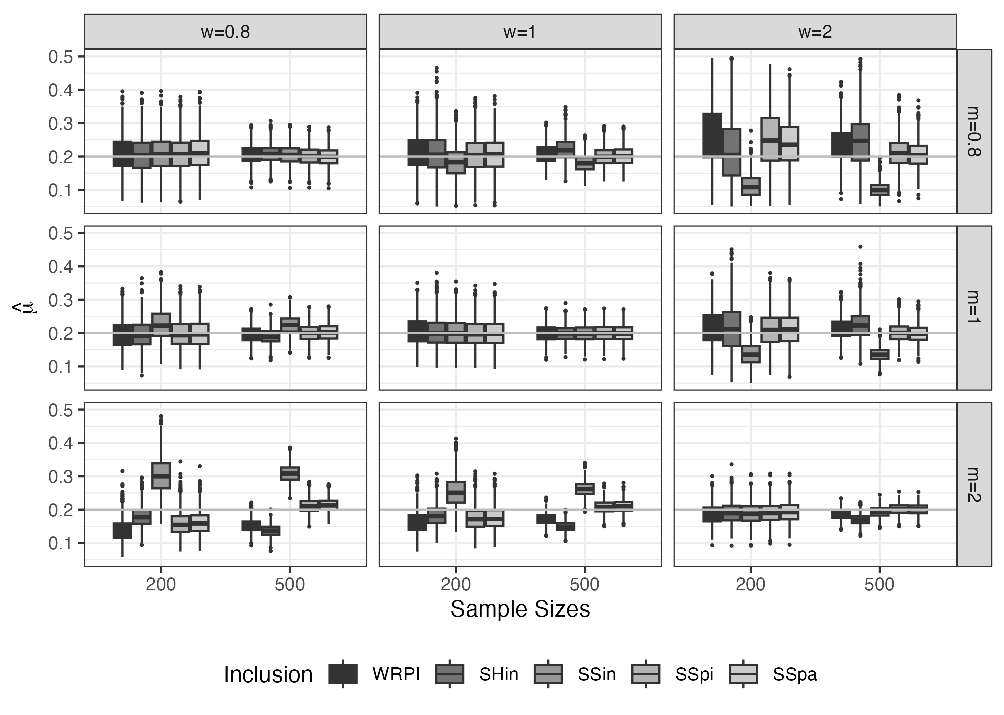}
\caption{Approxiated sampling distribution of the Hájek estimator based on WRPI, $SS_{in}$, $SS_{pi}$, $SS_{pa}$ and the $SH_{in}$ estimator when the network has different values of attractiveness $m$  and activity $w$ ratios, for samples of sizes 200 and 500. The homophily is high ($h=5$), and the proportion of directed edges is 0.2.}
\label{boxplot1}
\end{figure}

Figure \ref{boxplot1} shows the approximated sampling distribution of the prevalence estimators to help explain the findings presented in Figure~\ref{MSE_prev_h5}. In most scenarios, the differences in RSME appear primarily explained by the bias, with $SS_{pi}$ and $SS_{pa}$ exhibiting lower bias.

Overall, the Hájek estimators based on $SS_{pi}$ and $SS_{pa}$ result in comparable or better RSME than the alternative estimators in directed networks. They especially improve the inference compared to WRPI for scenarios with high homophily, when the attractiveness and activity ratios differ, and when the sample size is large. 

Also, they are preferred over $SS_{in}$ when the activity and attractiveness ratios are unequal. In that case, the gap is particularly significant for small sample sizes. The sample sizes 750 and 1125 results are shown in the supplementary material.

\section{Application}\label{sec:application}
To assess the performance of the proposed methods with real data, we use the `Wikipedia vote network' \footnote{Wikipedia is a free encyclopedia written collaboratively by volunteers worldwide.} data obtained from \cite{snapnets}. These data represent a directed network and contain all administrator elections and vote history. An administrator is a user with access to technical features that aid the platform's maintenance. The 4,159 nodes in the network are Wikipedia users, and a directed edge from node $i$ to node $j$ indicates that user $i$ voted for user $j$. There are 194,406 edges. We assigned the infection status to obtain a 20\% prevalence. The first 832 nodes in the dataset are assumed to be infected, and the remaining 3,327 nodes are uninfected. This assignment led to $h=4.6$, $m=1.19$, and $w=1.19$. We call this network $\text{Net}_1$. To assess alternative values of $m$ and $w$, we randomly removed $90\%$ and $70\%$ of the edges on the upper and lower triangles of the adjacency matrix for the infected nodes, respectively. We obtained two additional networks with $h=4.63$, $m=1.03$, $w=0.4$ and $h=4.65$, $m=0.58$, and $w=1.07$, respectively. We call these networks $\text{Net}_2$ and $\text{Net}_3$.

We drew 200 samples of size 1,386 from each network with the different sampling schemes: RDS, WRPI, $SS_{in}$, $SS_{pi}$, and $SS_{pa}$. We calculated the inclusion probability based on Equation (\ref{eq:pihat}), where $n_{samp} = 200$, and estimated the prevalence using those sampling probabilities. We compared the results across sampling approximations using the Mean Absolute Relative Error (MARE), described in Equation (\ref{MARE}).

Table \ref{tab:02} displays the MARE of the inclusion probability obtained by the sampling processes WRPI, $SS_{in}$, $SS_{pi}$, and $SS_{pa}$. As expected, when $m=w$ (network $\text{Net}_1$), the inclusion probabilities for the various RDS approximations are similar, except for WRPI, which is affected by the finite population effect. When $m\neq w$ (networks $\text{Net}_2$ and $\text{Net}_3$), the successive sampling with probability proportional to the partial in-degree ($SS_{pi}$) and its approximation ($SS_{pa}$) have the lowest MARE, about $30\%$ lower than $SS_{in}$ in some cases. These results are consistent with the simulation study.

\begin{figure}[!t]
\centering
\includegraphics[width = 11cm]{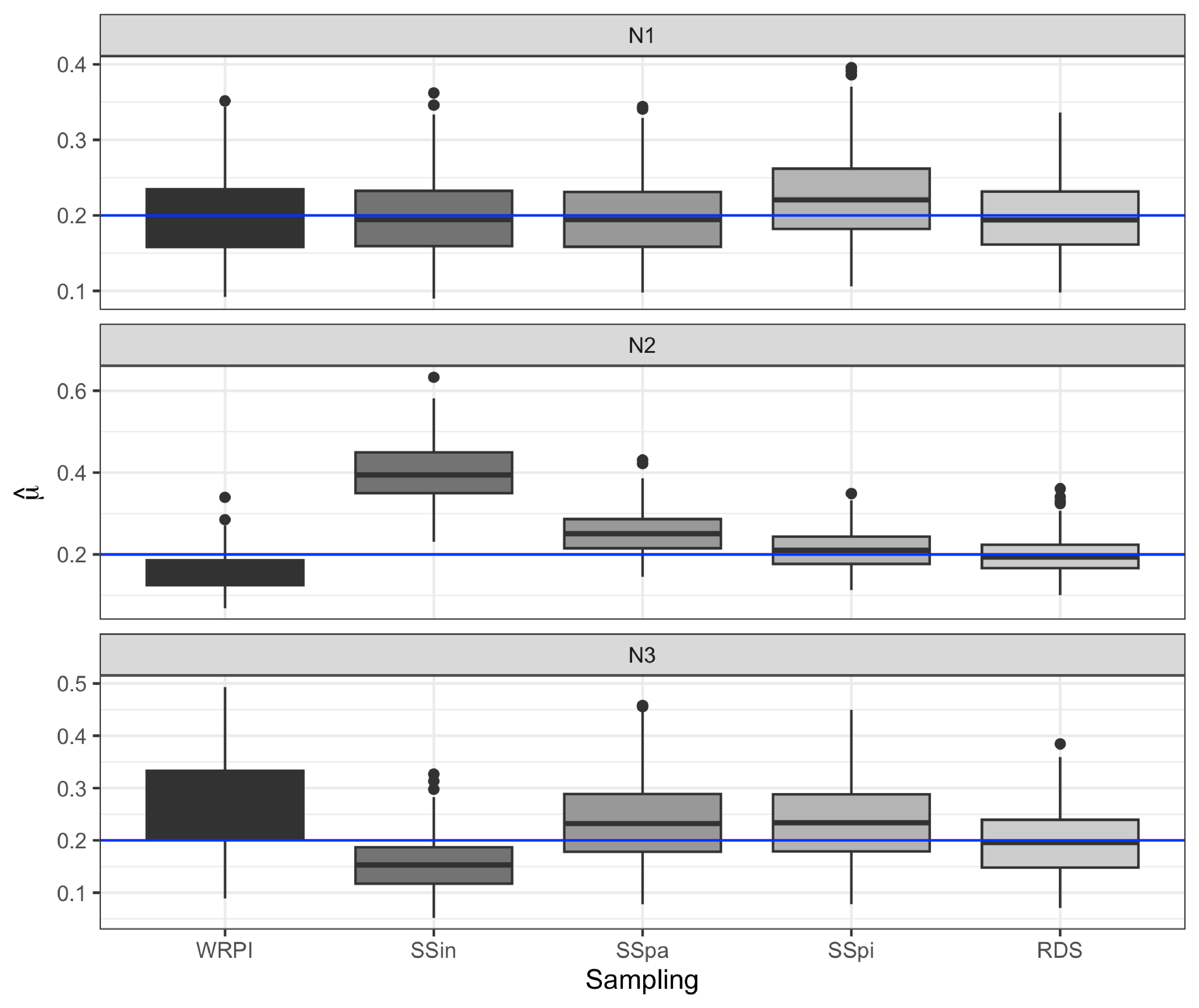}
\caption{Approximated sampling distribution of the Hájek estimator using the inclusion probabilities obtained by RDS, $SS_{in}$, $SS_{pi}$, $SS_{pa}$ and WRPI for the networks $\text{Net}_1, \text{Net}_2$ and $\text{Net}_3$ obtained from 'Wikipedia vote network'}
\label{figura_wiki_est}
\end{figure}

We also estimated the prevalence for each RDS sample using the Hájek estimators with the simulated inclusion probabilities. Figure \ref{figura_wiki_est} shows the results for each network and sampling design. The approximated sampling distributions are fairly similar when the attractiveness and activity ratios coincide ($m=w$). However, when they differ ($m\neq w$), the estimators based on WRPI and $SS_{in}$ are more biased than $SS_{pi}$ and $SS_{pa}$. Again, these results are consistent with the simulation study.

\begin{table}[!ht]
\caption{Mean absolute relative error (MARE) for the inclusion probabilities obtained by the different sampling processes: WRPI, $SS_{in}$, $SS_{pi}$,  and $SS_{pa}$. \label{tab:02}}
\begin{tabular*}{\columnwidth}{@{\extracolsep\fill}ccccc@{\extracolsep\fill}}
\toprule
 Network & WRPI & $SS_{in}$ & $SS_{pi}$ & $SS_{pa}$ \\ 
\midrule
$\text{Net}_1$ & 0.388 & 0.198 & 0.182 & 0.205  \\
$\text{Net}_2$ &  0.604 & 0.365 & 0.272 & 0.324 \\
$\text{Net}_3$ & 0.438 & 0.298 &0.192 & 0.213  \\
\toprule
\end{tabular*}
\end{table}

\section{Conclusions}\label{sec:conclusion}

Respondent-driven sampling (RDS) was first widely adopted in studies for hard-to-reach populations most vulnerable to HIV. In recent years, however, RDS has been employed in a much wider variety of applications. Although RDS provides a convenient solution to collecting data in socially connected populations, the validity of the inference highly depends on the network structure and participants' behaviours. This work's main contribution is to provide a model for the sampling probabilities when RDS is conducted in homophilous and partially directed networks. This overarching objective may be divided into two main contributions: First, we introduced a novel configuration model for directed and homophilous networks, and second, we proposed two approaches to approximate the RDS sampling probabilities using random walks over this configuration model.

Our network model, the attributed configuration model for a partially directed network (ACM), aims to capture the relationship between infected or uninfected nodes in partially directed networks. We provided the algorithm to generate networks from this model. Also, in the supplementary material, we proved that the resulting empirical degree distribution converges in probability to the desired target distribution.

We then derived a successive sampling approximation to RDS where the unit sizes are proportional to the partial in-degree ($SS_{pi}$). This representation leverages the information about the node characteristic so the recruitment patterns may accurately reflect the connectivity within and between the infection groups and the edges' direction. However, since the partial in-degrees are unlikely to be known in practice, we also proposed an RDS representation relying on approximating these quantities~($SS_{pa}$).

Through a simulation study, we compared the pseudo-inclusion probabilities from various RDS approximations for directed networks with the RDS inclusion probabilities. We drew samples of sizes 200, 500, 750, and 1125 from a simulated population of 1500 nodes. We showed that when the activity and attractiveness ratios are equal ($w = m$), the pseudo-inclusion probabilities from the SS methods are similar. However, when they diverge ($w\ne m$), the $SS_{pi}$ and $SS_{pa}$ are closer to the RDS inclusion probabilities compared to $SS_{in}$ and WRPI. This effect is exacerbated by the network homophily ($h = 5$). Also, as expected, the successive sampling approximations improve with a larger sample size, which is not the case for WRPI since it is affected by the finite population effect.

We subsequently used the inclusion probability approximations to estimate the prevalence of infected nodes using the Hájek estimator. We also compared the results with the $SH_{in}$ estimator designed for directed networks. As in the previous analysis, a high level of homophily and differences between attractiveness and activity ratio generate bias in the estimators based on the inclusion probabilities given by $SS_{in}$ or WRPI. The Hájek estimator based on WRPI and the $SH_{in}$ estimator display more considerable bias as the sample size increases. 

However, the Hájek estimators based on $SS_{pi}$ and $SS_{pa}$ improve the prevalence estimations in many of the considered scenarios. Our simulations have shown that incorporating infection status into the approximation of inclusion probabilities may reduce the estimators' bias and variance, even in the presence of homophily and when the attractiveness and activity ratios differ.

Finally, we applied the methodologies to the Wikipedia vote network and two modified versions where the edges were removed to obtain specific values of the parameters $m$ and $w$. The results were consistent with the simulation study. The approximations by $SS_{pi}$ or $SS_{pa}$ were closer to RDS, decreasing the mean squared error of the estimated inclusion probabilities and prevalence.

The findings from the simulation study and application suggest that the proposed approximations to RDS ($SS_{pi}$ or $SS_{pa}$) may significantly improve the estimation of the sampling probabilities in directed networks. This result is especially true in homophilous networks when the attractiveness and the activity ratios differ. This improvement also yields a more accurate prevalence estimation than the existing options for directed networks in those circumstances.
\singlespace
\bibliographystyle{plainnat}
\bibliography{ms}

\end{document}


\renewcommand{\thefootnote}{\fnsymbol{footnote}}
\renewcommand{\labelitemi}{$\diamond$}

\begin{center}
\textbf{\Large Supplementary Material: Partially Directed Configuration Model with Homophily and Respondent-Driven Sampling} \\

\vspace{1.5cc}

 \end{center}
\vspace{1 cc}

This is supplementary material to the main article. Section \ref{size_indegree} explains why using in-degree unit sizes is a suitable approximation for directed networks, as mentioned in Section 4 of the main article. Then, Section \ref{proof_lemma2} presents the proofs of lemmas 1 and 2 and the theorem 1. We also show the algorithm for generating the network in Section \ref{network_generation} and additional results of our simulations in Section \ref{more_results}.

\section{Successive Sampling proportional to in-degree ($SS_{in}$)}\label{size_indegree}

\cite{gile2011} considered a self-avoiding random walk over all possible undirected configuration networks where, at any given step, a node is randomly selected with probability proportional to its degree. The sampling scheme is equivalent to probability proportional to size without replacement sampling or successive sampling (SS). Then, \cite{xinlu2012} used two versions of SS for directed networks and compared the accuracy of the resulting prevalence estimates with their proposed estimator. One of the SS approaches sets the unit sizes as the out-degrees, and the other, the in-degrees. This section of the supplementary material justifies the appropriateness of the in-degree unit sizes for directed networks as we compare our results to this method.

The argument for setting the unit sizes equal to the in-degrees relies on the configuration network introduced by \cite{2015_britton_conf}. Their work extended the directed configuration model of \citeauthor{chen2013directed} to allow for partially directed edges. In particular, their partially directed configuration model (CMPD) is based on the distribution of the number of incoming stubs ($d^{\leftarrow}$), outgoing stubs ($d^{\rightarrow}$), and undirected stubs ($d^{\leftrightarrow}$), where stubs are unconnected half-edges. This network construction lets us incorporate both directed and undirected edges.

Under the CMPD, each node is initially assigned a fixed number of stubs (incoming, outgoing, and undirected). The directed stubs are joined randomly with another in the opposite direction to create a directed edge. Similarly, the undirected stubs are randomly connected with another undirected stub to create an undirected edge. After this process, loops are erased, and parallel edges are converted into undirected edges to make a simple graph. The resulting edge distribution may differ from the original stub distribution. However, \cite{2015_britton_conf} shows that the edge distribution converges to the stub distribution under some restrictions on the first moments of the stub distribution. The edge distribution is also referred to as the degree distribution.

If we assume RDS may be modelled using a self-avoiding random walk $G$ over all CMPD networks of fixed degree distribution, then node $g_{j-1}$ may visit node $g_j$ if node $g_{j-1}$ has an outgoing or undirected edge and node $g_j$ has an incoming or undirected edge. Consequently, the transition probability depends on incoming and undirected edges of node $g_j$ and is given by:

\begin{align}\label{conf_model_5}
  P(G_{j}=g_j|G_1,\ldots, G_{j-1}&=g_1,\ldots, g_{j-1})\\ \nonumber
  &=\begin{cases}
                                                    \frac{d^{in}_{g_j}}{E-\sum_{k=1}^{j-1} d_{g_{k}}^{in}} & g_j \notin g_1, \ldots, g_{j-1} \\
0, & g_j \in g_1, \ldots, g_{j-1}
                                                  \end{cases}
\end{align}
where $E=\sum_{j=1}^{N}d^{in}_{j}$ and $d^{in}_{j}=d_j^{\leftarrow}+d_{j}^{\leftrightarrow}$. 
This procedure is equivalent to sampling without replacement with probability proportional to in-degree. The algorithm proposed by \cite{gile2011} may then be used with the transition probabilities in Equation~(\ref{conf_model_5}) to estimate the pseudo-sampling probabilities $\pi$ with unit sizes equal to the in-degree.

\section{Proof of Lemma 1, Lemma 2 and Theorem 1} \label{proof_lemma2}
In this section, we prove Lemmas 1, 2 and Theorem 1 presented in Section 3.2 of the main article. The proofs follow the structure of \cite{2015_britton_conf} except that the variables and their distributions depend on the variable $Z$. This extension requires that we assume the ratio $\phi$ is fixed, and therefore, $N^1\to\infty$ and $N^0\to\infty$ as $N\to\infty$.

\subsection{Proof of Lemma 1}
The Lemma 1 says that $V^{(N^{z})}_{\boldsymbol{d}^{z}}/N^{z} \xrightarrow{P} p_{\boldsymbol{d}^{z}}$ implies that $F^{(N^{z})}_{\boldsymbol{D}^{z}}\to F_{\boldsymbol{D}^{z}}$ as $N^z\to\infty$. Similarly to \citeauthor{2015_britton_conf}, the proof is as follows:
\begin{enumerate}
\item[(a)] $V^{(N^z)}_{\boldsymbol{d}^z}/N^z \xrightarrow{P} p_{\boldsymbol{d}^z}$ and $0\le V^{(N^z)}_{\boldsymbol{d}^z}/N^z<1$ imply $E\left(V^{(N^z)}_{\boldsymbol{d}^z}/N^z\right) \rightarrow p_{\boldsymbol{d}^z}$ 
by bounded convergence \citep{2001_grimmett_probability}.
\item[(b)] $E\left(V^{(N^z)}_{\boldsymbol{d}^z}/N^z\right)=p^{(N^z)}_{\boldsymbol{d}^z}$ implies $p^{(N^z)}_{\boldsymbol{d}^z}\to p_{\boldsymbol{d}^z} \forall~\boldsymbol{d}^{z}$.
\item[(c)] Since (b) holds for any $\boldsymbol{d}^{z}$, we have that $F^{(N^{z})}_{\boldsymbol{D}^{z}}\to F_{\boldsymbol{D}^{z}}$ as $N^z\to \infty$. \hspace{1cm} \qedsymbol{}
\end{enumerate}

\subsection{Proof of Lemma 2}

The proof follows the one shown in Section 4 of \cite{2015_britton_conf}. Their construction is applicable under our conditions since we assume $\phi$ is constant. Analogue to their argument, we maintain that in order to establish that:
\begin{flalign}
\frac{V^{(N^{z})}_{\boldsymbol{d}^{z}}}{N^{z}}\xrightarrow{P} p_{\boldsymbol{d}^{z}},
\end{flalign}
as $N^{z}\to \infty$, it suffices to show that:
\begin{flalign}\label{eq:DeltaNodes}
\frac{\tilde{V}^{(N^{z})}_{\boldsymbol{d}^{z}}-V^{(N^{z})}_{\boldsymbol{d}^{z}}}{N^{z}}\xrightarrow{P} 0,
\end{flalign}
as $N^{z}\to \infty$, where $\tilde{V}^{(N^{z})}_{\boldsymbol{d}^{z}}$ is the number of nodes with degree $\boldsymbol{d}^{z}$ before making the network simple. 

Since the difference in the number of nodes with degree $\boldsymbol{d}^{z}$ before and after making the network simple has to be less than or equal to the total number of nodes for which a change has occurred, proving the following guarantees the convergence shown in Equation (\ref{eq:DeltaNodes}): 
\begin{flalign}
P(|M^{(N^{z})}/N^z|>\epsilon)\to 0 \text{ as } N^z \to\infty,~\forall~\epsilon >0,
\end{flalign}
where $M^{(N^{z})}=\sum_{r=1}^N M_r^{(N^{z_r})}\cdot\mathds{1}_{[z_{r}=z]}$ and $M_r^{(N^{z_r})}$ is equal to one if simplifying the network has changed the degree of node $r$, $\boldsymbol{d}^{z}_r$, and zero otherwise. As pointed out by \citeauthor{2015_britton_conf}, since $M^{(N^{z})}$ is a nonnegative random variable by construction, we can use Markov's inequalities to argue that:
\begin{flalign}
P(|M^{(N^{z})}/N^z|>\epsilon) \le \frac{E(M^{(N^{z})}/N^z)}{\epsilon}, ~\forall~\epsilon > 0.
\end{flalign}
Therefore, showing that $E(M^{(N^{z})}/N^z) \to 0$ would prove that $M^{(N^{z})}/N^z$ converges in probability to zero. To achieve this, the next step consists in showing that:
\begin{flalign}\label{eq:EMN}
E(M^{(N^{z})}/N^z) = P(M_r^{(N^{z_r})}=1)\to 0.
\end{flalign}
The equality in Equation (\ref{eq:EMN}) holds because, for $\{i: z_i = z_r\}$, 
$M_i^{(N^{z_i})}$ are identically distributed. 
Since $M^{(N^{z_r})}_{r}$ is either equal to one or zero, showing that $P(M_r^{(N^{z_r})}=1)\to 0 \text{ as } N^z \to\infty$ is equivalent to showing that $P(M_r^{(N^{z_r})}=0)\to 1$ as $N^z \to\infty$. 

Without loss of generality, we may fix $r=1$ and $z_1$. The event $M^{(N^{z_1})}_{1}= 0$ occurs when each stub from node $1$ is preserved after removing self-loops or parallel edges as described in Section 3.1 of the main article. The probability that it occurs may be written as follows:
\begin{flalign}
P(M^{(N^{z_1})}_{1} =0) = \sum_{\boldsymbol{d}_1^{z_1}} P(M^{(N^{z_1})}_{1}=0|\boldsymbol{D}_1^{z_1}= \boldsymbol{d}_1^{z_1}) P(\boldsymbol{D}_1^{z_1}= \boldsymbol{d}_1^{z_1}),
\end{flalign}
where $\boldsymbol{d}_1^{z_1} = (d^{z_1\leftarrow 1}_1,d^{z_1\rightarrow 1}_1,d^{z_1\leftrightarrow 1}_1,d^{z_1\leftarrow 0}_1,d^{z_1\rightarrow 0}_1, d^{z_1\leftrightarrow 0}_1)$. Therefore, if
\begin{flalign}
P(M^{(N^{z_1})}_{1}|\boldsymbol{D}_1^{z_1} = \boldsymbol{d}_1^{z_1}) \to 1~\forall~\boldsymbol{d}_1^{z_1} \text{ as } N^{z_1}\to\infty,
\end{flalign}
then $P(M_r^{(N^{z_r})}=0)\to 1 \text{ as } N^z \to\infty$. \hspace{6cm} \qedsymbol{}

\subsection{Proof of Theorem 1}

In this section, we prove Theorem~1 following closely the arguments presented in \cite{2015_britton_conf}, which we adapt to include the infection status of the nodes reflected in the ACM. Theorem 1 states that, under certain conditions, (a)~the empirical degree distribution of the ACM converges to the target degree distribution of the ACM and (b)~the empirical proportion of nodes in the ACM with a given degree converges in probability to the target proportion of that degree in the ACM. As established by Lemma~1, it suffices to demonstrate~(b) to prove Theorem~1 as (b)~implies (a). Also, Lemma~2 states that, to prove~(b), it suffices to show that $P(M_1^{(N^{z_1})}=0 \mid \boldsymbol{D}_1^{z_1} = \boldsymbol{d}_1^{z_1})\to 1, \forall~\boldsymbol{d}_1^{z_1}$ as $N^{z_1}\to\infty$. We first note that:
\begin{flalign}
    P(M_1^{(N^{z_1})}=0 \mid \boldsymbol{D}_1^{z_1} = \boldsymbol{d}_1^{z_1}) = 
    E[P(M_1^{(N^{z_1})}=0 \mid \boldsymbol{D}_1^{z_1} = \boldsymbol{d}_1^{z_1},\boldsymbol{D}^{N})],
\end{flalign}
where $\boldsymbol{D}^{N} = (\boldsymbol{D}_2^{z_2},\ldots ,\boldsymbol{D}_N^{z_N})$. Consequently, our objective is to demonstrate that:
\begin{flalign}\label{eq:EPM1eq0}
    E[P(M_1^{(N^{z_1})}=0 \mid \boldsymbol{D}_1^{z_1} = \boldsymbol{d}_1^{z_1},\boldsymbol{D}^{N})] \to 1, \forall~\boldsymbol{d}_1^{z_1} \text{ as } N^{z_1}\to\infty.
\end{flalign}
We start by discussing $P(M_1^{(N^{z_1})}=0 \mid \boldsymbol{D}_1^{z_1} = \boldsymbol{d}_1^{z_1},\boldsymbol{D}^{N})$, and then determine its expected value. This statement may be interpreted as the probability that the degree of node one is not modified in the process of making the network simple conditional on the degree distribution, where the degree of node one is 
$\boldsymbol{d}^{z_1}_1=(d_1^{z_1\leftarrow 1},d_1^{z_1\leftarrow 0}, d_1^{z_1\rightarrow 1},d_1^{z_1\rightarrow 0}, d_1^{z_1 \leftrightarrow 1},d_1^{z_1\leftrightarrow 0})$. 

In order for that degree to remain unchanged, each of the six components needs to stay the same. Let us denote $q^{z_1\leftarrow 1}_{\boldsymbol{i}}$ the event that the first component of the degree $d_1^{z_1\leftarrow 1}$ has not changed while attaching to nodes with indices $\boldsymbol{i} = (i_1,i_2,\ldots,i_{d^{z_1\leftarrow 1}})$ conditional on $\boldsymbol{D}_1^{z_1}$ and $\boldsymbol{D}^{N}$. Since we know the entire degree distribution, we can calculate the probability that this event occurs as follows:
\begin{align}\label{eq:qz11}
q^{z_1\leftarrow 1}_{\boldsymbol{i}} =&\prod_{r=1}^{d^{z_1\leftarrow 1}_1} P(\text{node } i_r \text{ connects to node 1 with one of its ``} 1\rightarrow z_1 \text{'' stubs}) ,\notag\\
=&\prod_{r=1}^{d^{z_1\leftarrow 1}_1}
\frac{i_r \text{'s number of ``} 1\rightarrow z_1 \text{'' stubs}}{
\text{total number of ``} 1\rightarrow z_1 \text{'' stubs available}
},\notag\\
=&\prod_{r=1}^{d^{z_1\leftarrow 1}_1 }
\frac{d_{i_r}^{1\rightarrow z_1}\mathds{1}_{[z_{i_r}=1]}} 
{s^{1 \rightarrow z_1}-r+1+u^{z_1\leftarrow 1}},
\end{align}
where:
\begin{enumerate}
    \item[(a)] $s^{1 \rightarrow z_1} = \sum_{i = 1}^N d_i^{1\rightarrow z_1}\mathds{1}_{[z_{i}=1]}$,
    \item[(b)] $u^{z_1\leftarrow 1}=d^{z_1 \leftarrow 1}_1 \mathds{1}_{[d^{z_1\leftarrow 1}_1>s^{1 \rightarrow z_1}]}+\omega^{(N)}_{(z_1,{}_1)}\mathds{1}_{[\omega^{(N)}_{(z_1,{}_1)}>0]}$,
    \item[(c)] $\omega^{(N)}_{(z_1,{}_1)} = s^{z_1\leftarrow 1}-s^{1\rightarrow z_1}$.
\end{enumerate}
In the denominator of Equation (\ref{eq:qz11}), the term $s^{1 \rightarrow z_1}$ counts the number of outgoing stubs of type ``$1\rightarrow z_1$'' in the network. The total $s^{1 \rightarrow z_1}$ is reduced by $r-1$ to take into account that $r-1$ stubs are no longer available at the time node $i_r$ is attempting to connect one of its ``$1\rightarrow z_1$'' stubs to node 1. For instance, $i_2$ is the second node to try to connect to node one. At the time of the connection, the total number of ``$1\rightarrow z_1$ '' stubs to choose from is $s^{1 \rightarrow z_1} - (2-1) = s^{1 \rightarrow z_1} - 1$ since we have already used one stub of type ``$z_1\leftarrow 1$'' when connecting node $i_1$ to node one. As for the term $u^{z_1\leftarrow 1}$, it ensures that the denominator is positive. There are two instances where this may not happen:~(1) if the degree $d_1^{z_1\leftarrow 1}$ is greater than the number of ``$1\rightarrow z_1$'' stubs and (2)~when the number of ``$z_1\leftarrow 1$'' stubs is greater than the number of ``$1\rightarrow z_1$'' stubs.

In order to calculate the expected value in Equation ($\ref{eq:EPM1eq0}$), we rewrite $q^{z_1\leftarrow 1}_{\boldsymbol{i}}$ as a random variable $Q^{z_1\leftarrow 1}_{\boldsymbol{i}}$ which depends on $\boldsymbol{D}^{N}$ so that:
\begin{align}\label{eq:Qi}
Q^{z_1\leftarrow 1}_{\boldsymbol{i}} =&\prod_{i_r=1}^{d^{z_1\leftarrow 1}_1}
\frac{D_{i_r}^{1\rightarrow z_1}\mathds{1}_{[z_{i_r}=1]}}
{S^{1 \rightarrow z_1}-r+1+U^{z_1\leftarrow 1}},
\end{align}
where the uppercase letters indicate the associated random variables. So far, we have focused only on the first component of the $\boldsymbol{d}^{z_1}_1$ vector, that is, on $d_1^{z_1\leftarrow 1}$. More specifically, we have derived the probability that $d_1^{z_1\leftarrow 1}$ would not be changed while attaching to nodes with indices $\boldsymbol{i} = (i_1,i_2,\ldots,i_{d^{z_1\leftarrow 1}})$ conditional on $\boldsymbol{D}_1^{z_1}$ and $\boldsymbol{D}^{N}$. To determine the probability that none of the six components would be altered conditional on $\boldsymbol{D}_1^{z_1}$ and $\boldsymbol{D}^{N}$, $P(M_1^{(N^{z_1})}=0 \mid \boldsymbol{D}_1^{z_1} = \boldsymbol{d}_1^{z_1},\boldsymbol{D}^{N})$, we have to consider all six components and for all possible selection of indices node one might connect to. In other words:
\begin{align}
P(&M^{(N^{z_1})}_{1}= 0|\boldsymbol{D}_1^{z_1},\boldsymbol{D}^{N})= \nonumber\\
&\sum_{\boldsymbol{i},\boldsymbol{j},\boldsymbol{k},\boldsymbol{l},\boldsymbol{m},\boldsymbol{o}}
Q^{z_1\leftarrow 1}_{\boldsymbol{i}}
Q^{z_1\leftarrow 0}_{\boldsymbol{j}}
Q^{z_1\rightarrow 1}_{\boldsymbol{k}}
Q^{z_1\rightarrow 0}_{\boldsymbol{l}}
Q^{z_1\leftrightarrow 1}_{\boldsymbol{m}}
Q^{z_1\leftrightarrow 0}_{\boldsymbol{o}},\label{eq_suma2}
\end{align}
where the terms 
$Q^{z_1\leftarrow 1}_{\boldsymbol{i}},
Q^{z_1\leftarrow 0}_{\boldsymbol{j}},
Q^{z_1\rightarrow 1}_{\boldsymbol{k}},
Q^{z_1\rightarrow 0}_{\boldsymbol{l}},
Q^{z_1\leftrightarrow 1}_{\boldsymbol{m}},$ and 
$Q^{z_1\leftrightarrow 0}_{\boldsymbol{o}}$ are defined in Table~\ref{tab:Qs1} and \ref{tab:Qs2}. Equation (\ref{eq_suma2}) is taking the sum over all values of 
$\boldsymbol{i} = (i_1,i_2,\ldots,i_{d^{z_1\leftarrow 1}}) $,\\
$\boldsymbol{j} = (j_1,j_2,\ldots,j_{d^{z_1\leftarrow 0}})$, 
$\boldsymbol{k} = (k_1,k_2,\ldots,k_{d^{z_1\rightarrow 1}})$, 
$\boldsymbol{l} = (l_1,l_2,\ldots,l_{d^{z_1\rightarrow 0}})$, 
$\boldsymbol{m} = (m_1,m_2,$
\\
$\ldots,m_{d^{z_1\leftrightarrow 1}})$, and 
$\boldsymbol{o} = (o_1,n_2,\ldots,o_{d^{z_1\leftrightarrow 0}})$, such that none of the indices are equal to one (to avoid self-loops) and all indices are unique, that is, the stubs of node one are all pointing to distinct nodes. If it were not the case, the degree of node~one would be altered when making the network simple.

As in Equation (\ref{eq:Qi}), adjustments to the denominators of 
$Q^{z_1\leftarrow 0}_{\boldsymbol{j}},
Q^{z_1\rightarrow 1}_{\boldsymbol{k}}$,
$Q^{z_1\rightarrow 0}_{\boldsymbol{l}},$\\
$Q^{z_1\leftrightarrow 1}_{\boldsymbol{m}},$ and 
$Q^{z_1\leftrightarrow 0}_{\boldsymbol{o}}$
are included through the variables $U$. Besides ensuring the denominators are strictly positive, the number of incoming stubs in the denominators of $Q^{z_1\rightarrow 1}_{\boldsymbol{k}}$, and 
$Q^{z_1\rightarrow 0}_{\boldsymbol{l}}$ is further reduced 
by $d_{1}^{1\leftarrow z_1}$ and $d_{1}^{0 \leftarrow z_1}$, respectively. This additional adjustment takes into account that some stubs from node one have already been attached to other nodes in $Q^{z_1\leftarrow 1}_{\boldsymbol{i}}$
and $Q^{z_1\leftarrow 0}_{\boldsymbol{j}}$, respectively. Finally, a factor of 2 is introduced in the denominators of $Q^{z_1\leftrightarrow 1}_{\boldsymbol{m}}$ and $Q^{z_1\leftrightarrow 0}_{\boldsymbol{o}}$ for 
reciprocal stubs sharing the same infection status. Those factors ensure that the total number of stubs available is reduced by the two reciprocal stubs of the same type used when a connection is formed.

We note that Equation (\ref{eq_suma2}) may include the summation of zero terms if at least one index in $\boldsymbol{i}, \boldsymbol{j}, \boldsymbol{k}, \boldsymbol{l}, \boldsymbol{m}$ or $\boldsymbol{o}$ cannot connect to node 1. For instance, if  $d_{i_r}^{1\rightarrow z_1} = 0$ or if $z_{i_r} =0$, then node 1 is not able to connect its $z_1 \rightarrow 1$ stubs to node $i_r$. In such a case, $Q^{z_1\leftarrow 1}_{\boldsymbol{i}}$ is equal to zero, and this case does not contribute to the probability that the degree remains unchanged.

\begin{landscape}
\begin{table}
\caption{Definition of the terms of Equation (\ref{eq_suma2})}

\footnotesize
\begin{tabular}{llp{0.75\textwidth}}
\hline\\&&\\[-18pt]
$Q^{z_1\leftarrow 1}_{\boldsymbol{i}} =\prod_{r=1}^{d^{z_1\leftarrow 1}_1}\frac{D_{i_r}^{1\rightarrow z_1}\mathds{1}_{[z_{i_r}=1]}}
{S^{1\rightarrow z_1}-r+1+U^{z_1\leftarrow 1}}$   & &
\vspace{-18pt}
\begin{enumerate}
    \item[]    $S^{1 \rightarrow z_1} = \sum_{i = 1}^N D_i^{1\rightarrow z_1}\mathds{1}_{[z_{i}=1]}$
    \item[] $U^{z_1\leftarrow 1}=d^{z_1 \leftarrow 1}_1 \mathds{1}_{[d^{z_1\leftarrow 1}_1>S^{1 \rightarrow z_1}]}+\Omega^{(N)}_{(z_1,{}_1)}\mathds{1}_{[\Omega^{(N)}_{(z_1,{}_1)}>0]}$
    \item[] $\Omega^{(N)}_{(z_1,{}_1)} = S^{z_1\leftarrow 1}-S^{1\rightarrow z_1}$
\end{enumerate} \\\hline\\&&\\[-18pt]
$Q^{z_1\leftarrow 0}_{\boldsymbol{j}}=\prod_{r=1}^{d^{z_1\leftarrow 0}_1}
\frac{D_{j_r}^{0\rightarrow z_1}\mathds{1}_{[z_{j_r}=0]}}
{S^{0 \rightarrow z_1}-r+1+U^{z_1\leftarrow 0}}$\\
& &\vspace{-30pt}
\begin{enumerate}
    \item[] $S^{0 \rightarrow z_1} = \sum_{i = 1}^N D_i^{0\rightarrow z_1}\mathds{1}_{[z_{i}=0]}$
    \item[] $U^{z_1\leftarrow 0}=d^{z_1\leftarrow 0}_1\mathds{1}_{[d^{z_1\leftarrow 0}_1>S^{0\rightarrow z_1}]}+\Omega^{(N)}_{(z_1,{}_0)}\mathds{1}_{[\Omega^{(N)}_{(z_1,{}_0)}>0]}$
    \item[] $\Omega^{(N)}_{(z_1,{}_0)} = S^{z_1\leftarrow 0}-S^{0\rightarrow z_1}$
\end{enumerate} \\\hline\\&&\\[-18pt]
$Q^{z_1 \rightarrow 1}_{\boldsymbol{k}}=\prod_{r=1}^{d^{z_1\rightarrow 1}_1}
\frac{D_{k_r}^{1\leftarrow z_1}\mathds{1}_{[z_{k_r}=1]}}
{S^{1 \leftarrow z_1}-d^{1\leftarrow z_1}_1\mathds{1}_{[z_1=1]}-r+1+U^{z_1 \rightarrow 1}}$  && \vspace{-18pt}
\begin{itemize}
    \item[] $S^{1 \leftarrow z_1} = \sum_{i = 1}^N D_i^{1 \leftarrow z_1}\mathds{1}_{[z_{i}=1]}$
    \item[] $U^{z_1 \rightarrow 1}=(
d^{z_1\rightarrow 1}_1 + d^{1\leftarrow z_1}_1\mathds{1}_{[z_1=1]})\mathds{1}_{[d^{z_1\rightarrow 1}_1>S^{1 \leftarrow z_1}-d^{1\leftarrow z_1}_1\mathds{1}_{[z_1=1]}]}$
    \item[] \hspace{1.35cm}$-~\Omega^{(N)}_{(z_1,{}_1)}\mathds{1}_{[\Omega^{(N)}_{(z_1,{}_1)}<0]}$ 
\end{itemize} \\\hline\\&&\\[-18pt]
    \end{tabular}
    \label{tab:Qs1}
\end{table}
\end{landscape}

\begin{landscape}
\begin{table}
\caption{Continuation of definition of the terms of Equation (\ref{eq_suma2})}
\begin{tabular}{llp{0.75\textwidth}}
\hline\\&&\\[-18pt]
$Q^{z_1\rightarrow 0}_{\boldsymbol{l}}=\prod_{r=1}^{d^{z_1\rightarrow 0}_1}
\frac{D_{l_r}^{0 \leftarrow z_1}\mathds{1}_{[z_{l_r}=0]}}
{S^{0 \leftarrow z_1}-d^{0\leftarrow z_1}_1\mathds{1}_{[z_1=0]}-r+1+U^{z_1 \rightarrow 0}}$&& 
\vspace{-18pt}
\begin{itemize}
    \item[] $S^{0 \leftarrow z_1} = \sum_{i = 1}^N D_i^{0\leftarrow z_1}\mathds{1}_{[z_{i}=0]}$
    \item[] $U^{z_1 \rightarrow 0}=(d^{z_1\rightarrow 0}_1+d^{0\leftarrow z_1}_1\mathds{1}_{[z_1=0]})\mathds{1}_{[d^{z_1\rightarrow 0}_1>S^{0 \leftarrow z_1}-d^{0\leftarrow z_1}_1\mathds{1}_{[z_1=0]}]}$
    \item[] \hspace{1.35cm}$ -~\Omega^{(N)}_{(z_1,{}_0)}\mathds{1}_{[\Omega^{(N)}_{(z_1,{}_0)}<0]}$
\end{itemize} \\\hline\\&&\\[-18pt]
$Q^{z_1\leftrightarrow 1}_{\boldsymbol{m}} =   \prod_{r=1}^{d^{z_1\leftrightarrow 1}_1}
\frac{D_{m_r}^{1\leftrightarrow z_1}\mathds{1}_{[z_{m_r}=1]}}
{S^{1 \leftrightarrow z_1}-2^{z_1}r+1+U^{z_1 \leftrightarrow 1}}$\\
&&\vspace{-30pt}
\begin{itemize}
    \item[] $S^{1 \leftrightarrow z_1} = \sum_{i = 1}^N D_i^{1 \leftrightarrow z_1}\mathds{1}_{[z_{i}=1]}$
    \item[] $U^{z_1 \leftrightarrow 1}=2^{z_1}d^{z_1\leftrightarrow 1}_1\mathds{1}_{[2^{z_1}d^{z_1 \leftrightarrow 1}_1>S^{1 \leftrightarrow z_1}]} + \Gamma^{(N)}_{(z_1,1)}\mathds{1}_{[\Gamma^{(N)}_{(z_1,1)}>0]}$
    \item[] $\Gamma^{(N)}_{(z_1,1)}=S^{z_1\leftrightarrow 1} - S^{1\leftrightarrow z_1}$
\end{itemize} \\\hline\\&&\\[-18pt]
$Q^{z_1\leftrightarrow 0}_{\boldsymbol{o}} =\prod_{r=1}^{d^{z_1\leftrightarrow 0}}
\frac{D_{o_r}^{0\leftrightarrow z_1}\mathds{1}_{[z_{n_r}=0]}}
{S^{0 \leftrightarrow z_1}
-d^{0 \leftrightarrow z_1}_1\mathds{1}_{[z_1=1]}
-2^{(1-z_1)}r+1+U^{z_1 \leftrightarrow 0}}$
&&\vspace{-18pt}
\begin{itemize}
    \item[] $S^{0 \leftrightarrow z_1} = \sum_{i = 1}^N D_i^{0 \leftrightarrow z_1}\mathds{1}_{[z_{i}=0]}$
    \item[] $U^{z_1 \leftrightarrow 0}= \Gamma^{(N)}_{(z_1,0)}\mathds{1}_{[\Gamma^{(N)}_{(z_1,0)}>0]}+ \Theta_{(z_1,0)} \mathds{1}_{[\Theta_{(z_1,0)}>S^{0 \leftrightarrow z_1} ]}$
    \item[] $\Theta_{(z_1,0)} = 2^{(1-z_1)}d^{z_1\leftrightarrow 0}_1 + d^{0 \leftrightarrow z_1}_1$
    \item[] $\Gamma^{(N)}_{(z_1,0)}=S^{z_1\leftrightarrow 0} - S^{0\leftrightarrow z_1}$
\end{itemize} \\\hline
    \end{tabular}
    \label{tab:Qs2}
\end{table}
\end{landscape}
\normalsize 

The next step consists of taking the expected value of Equation (\ref{eq_suma2}) to demonstrate the result in Equation (\ref{eq:EPM1eq0}). The expected value is as follows:
\begin{flalign}
&P(M^{(N^{z_1})}_{1}= 0|\boldsymbol{D}_1=\boldsymbol{d}_1)=E[P(M^{(N)}_{1}= 0|\boldsymbol{D}_1=\boldsymbol{d}_1,\boldsymbol{D}^{N}=\boldsymbol{d}^{N})] \nonumber\\
&=E\left[\sum_{\boldsymbol{i},\boldsymbol{j},\boldsymbol{j},\boldsymbol{l},\boldsymbol{m},\boldsymbol{o}}
Q^{z_1\leftarrow 1}_{\boldsymbol{i}}Q^{z_1\leftarrow 0}_{\boldsymbol{j}}Q^{z_1\rightarrow 1}_{\boldsymbol{k}}Q^{z_1\rightarrow 0}_{\boldsymbol{l}}Q^{z_1\leftrightarrow 1}_{\boldsymbol{m}}Q^{z_1\leftrightarrow 0}_{\boldsymbol{o}}\right]\nonumber\\
&=\sum_{\boldsymbol{i},\boldsymbol{j},\boldsymbol{j},\boldsymbol{l},\boldsymbol{m},\boldsymbol{o}}E\left[
Q^{z_1\leftarrow 1}_{\boldsymbol{i}}Q^{z_1\leftarrow 0}_{\boldsymbol{j}}Q^{z_1\rightarrow 1}_{\boldsymbol{k}}Q^{z_1\rightarrow 0}_{\boldsymbol{l}}Q^{z_1\leftrightarrow 1}_{\boldsymbol{m}}Q^{z_1\leftrightarrow 0}_{\boldsymbol{o}}
\right]\nonumber\\
&=c\cdot E\left[(N^1)^{d^{(z_1,1)}_1}(N^0)^{d^{(z_1,0)}_1}
Q^{z_1\leftarrow 1}_{\boldsymbol{i}}
Q^{z_1\leftarrow 0}_{\boldsymbol{j}}Q^{z_1\rightarrow 1}_{\boldsymbol{k}}
Q^{z_1\rightarrow 0}_{\boldsymbol{l}}Q^{z_1\leftrightarrow 1}_{\boldsymbol{m}}
Q^{z_1\leftrightarrow 0}_{\boldsymbol{o}}\right]\nonumber\\
&=c\cdot E\left[
\left\{(N^1)^{d^{z_1\leftarrow 1}_1}
Q^{z_1\leftarrow 1}_{\boldsymbol{i}}\right\}
\left\{(N^1)^{d^{z_1\rightarrow 1}_1}
Q^{z_1\rightarrow 1}_{\boldsymbol{k}}\right\}
\left\{(N^1)^{d^{z_1\leftrightarrow 1}_1}
Q^{z_1\leftrightarrow 1}_{\boldsymbol{m}}\right\} \right. \nonumber\\
&\hspace{1.4cm}\left. 
\left\{(N^0)^{d^{z_1\leftarrow 0}_1}
Q^{z_1\leftarrow 0}_{\boldsymbol{j}}\right\}
\left\{(N^0)^{d^{z_1\rightarrow 0}_1}
Q^{z_1\rightarrow 0}_{\boldsymbol{l}} \right\}
\left\{(N^0)^{d^{z_1\leftrightarrow 0}_1}
Q^{z_1\leftrightarrow 0}_{\boldsymbol{o}}\right\}\right]
\label{eq_lim}
\end{flalign}
where:
\begin{enumerate}\setlength\itemsep{6pt}
    \item[(a)] $d^{(z_1,1)}_1 = d^{z_1\leftarrow 1}_1+d^{z_1\rightarrow 1}_1+d^{z_1\leftrightarrow 1}_1$: node 1's stubs shared with infected nodes,
    \item[(b)] $d_1^{(z_1,0)} = d^{z_1\leftarrow 0}_1+d^{z_1\rightarrow 0}_1+d^{z_1\leftrightarrow 0}_1$: node 1's stubs shared with uninfected nodes,
    \item[(c)] $c=\frac{\displaystyle\prod_{v=0}^{d^{(z_{1},{}_1)}_1-1}(N^1-\mathds{1}_{[z_{i}=1]}-v)\cdot \prod_{v=0}^{d^{(z_1,0)}_1-1}(N^0-\mathds{1}_{[z_{i}=0]}-v)}{(N^{1})^{d^{(z_1,{}_1)}}(N^{0})^{d^{(z_{1},0)}}}.$
\end{enumerate}

\noindent To conclude the proof of Theorem 1, we need Lemma 3, which states that if \\(1)~$\lim_{m\to\infty} E[X_m]\leq a$, (2)~$X_m \xrightarrow{D} X$ as $m\to \infty$, and (3)~$E[X] = a$ then: $$\lim_{m\to \infty} E[X_m] = a,$$ 
where $\{X_m\}$ is a sequence of non-negative random variables, $X$ is a non-negative random variable and $0<a<\infty$ is a real number.\\

\noindent If we let $a = 1$ and let $\{X_{N}\}$ be a sequence of non-negative random variables such that:
\begin{flalign}
X_{N} &=
\left[(N^1)^{d^{z_1\leftarrow 1}_1}
Q^{z_1\leftarrow 1}_{\boldsymbol{i}}\right]
\left[(N^1)^{d^{z_1\rightarrow 1}_1}
Q^{z_1\rightarrow 1}_{\boldsymbol{k}}\right]
\left[(N^1)^{d^{z_1\leftrightarrow 1}_1}
Q^{z_1\leftrightarrow 1}_{\boldsymbol{m}}\right] \nonumber\\
&\hspace{1.4cm} 
\left[(N^0)^{d^{z_1\leftarrow 0}_1}
Q^{z_1\leftarrow 0}_{\boldsymbol{j}}\right]
\left[(N^0)^{d^{z_1\rightarrow 0}_1}
Q^{z_1\rightarrow 0}_{\boldsymbol{l}} \right]
\left[(N^0)^{d^{z_1\leftrightarrow 0}_1}
Q^{z_1\leftrightarrow 0}_{\boldsymbol{o}}\right]
\end{flalign}
then we have that the first condition of Lemma 3 is met, that is, $\lim_{N\to\infty} E[X_N]\leq 1$, since $P(M^{(N^{z_1})}_{1}= 0|\boldsymbol{D}_1=\boldsymbol{d}_1) \leq 1$ and $\lim_{N\to\infty} c = 1$. To show the second condition, we first rewrite the $X_N$ variable as follows: 
\begin{flalign}
X_N &=X_{N}^{\boldsymbol{i}}
X_{N}^{\boldsymbol{j}}
X_{N}^{\boldsymbol{k}}
X_{N}^{\boldsymbol{l}}
X_{N}^{\boldsymbol{m}}
X_{N}^{\boldsymbol{o}},
\end{flalign}
where:
\begin{align}
X_{N}^{\boldsymbol{i}} &= (N^1)^{d^{z_1\leftarrow 1}_1}
Q^{z_1\leftarrow 1}_{\boldsymbol{i}},\\
X_{N}^{\boldsymbol{j}} &= (N^0)^{d^{z_1\leftarrow 0}_1}
Q^{z_1\leftarrow 0}_{\boldsymbol{j}},\\
X_{N}^{\boldsymbol{k}} &= (N^1)^{d^{z_1\rightarrow 1}_1}
Q^{z_1\rightarrow 1}_{\boldsymbol{k}},\\
X_{N}^{\boldsymbol{l}}&=(N^0)^{d^{z_1\rightarrow 0}_1}
Q^{z_1\rightarrow 0}_{\boldsymbol{l}},\\
X_{N}^{\boldsymbol{m}} &= (N^1)^{d^{z_1\leftrightarrow 1}_1}
Q^{z_1\leftrightarrow 1}_{\boldsymbol{m}},\\
X_{N}^{\boldsymbol{o}} &= (N^0)^{d^{z_1\leftrightarrow 0}_1}
Q^{z_1\leftrightarrow 0}_{\boldsymbol{o}}.
\end{align}
If we take the limit of the $X_{N}^{\boldsymbol{i}}$ when $N\to\infty$ we get that:
\begin{align}
&\lim_{N\to \infty} X_N^{\boldsymbol{i}}=\notag\\
&\lim_{N\to \infty} \prod_{r=1}^{d^{z_1\leftarrow 1}_1}\frac{D_{i_r}^{1\rightarrow z_1}\mathds{1}_{[z_{i_r}=1]}}
{\frac{S^{1 \rightarrow z_1}}{N^{1}}-\frac{r+1}{N^1}+\frac{d^{z_1\leftarrow 1}_1}{N^1} \mathds{1}_{[d^{z_1\leftarrow 1}_1>S^{1 \leftarrow z_1}]}+\frac{\Omega^{(N)}_{(z_1, 1)}}{N^1}\mathds{1}_{[\Omega^{(N)}_{(z_1, 1)}>0]}}.
\end{align}
By the law of large numbers, we have that:
\begin{flalign}
\lim_{N\to\infty}\frac{\Omega^{(N)}_{(z_1,1)}}{N^{1}} &= \lim_{N\to \infty}\frac{S^{z_1\leftarrow 1}-S^{1\rightarrow z_1}}{N^{1}}
=\lim_{N\to \infty}\frac{N^{z_1}}{N^{1}}\frac{S^{z_1\leftarrow 1}}{N^{z_1}}-\frac{S^{1\rightarrow z_1}}{N^{1}}. 
\end{flalign}

\noindent For $z_1=0$, we have that:
\begin{flalign}
\lim_{N\to\infty}\frac{\Omega^{(N)}_{(0,1)}}{N^{1}} &=\lim_{N\to \infty}\frac{1}{\phi}\frac{S^{0\leftarrow 1}}{N^{0}}-\frac{S^{1\rightarrow 0}}{N^{1}} \notag \\
&=\frac{1}{\phi}\delta^{0\leftarrow 1}-\delta^{1 \rightarrow 0}=0, \quad \text{(by assumption)}
\end{flalign}

\noindent and for $z_1=1$, we have that:
\begin{flalign}
\lim_{N\to\infty}\frac{\Omega^{(N)}_{(1,1)}}{N^{1}} &=\lim_{N\to \infty}\frac{S^{1\leftarrow 1}}{N^{1}}-\frac{S^{1\rightarrow 1}}{N^{1}} \notag \\
&=\delta^{1\leftarrow 1}-\delta^{1 \rightarrow 1}=0 \quad \text{(by assumption)}.
\end{flalign}
Therefore, this leads to: \quad 

$X_N^{\boldsymbol{i}}\xrightarrow{D}
\frac{\prod_{r=1}^{d^{z_1\leftarrow 1}_1}D_{i_r}^{1\rightarrow z_1}\mathds{1}_{[z_{i_r}=1]}}{(\delta^{1\leftarrow z_1})^{d^{z_1\leftarrow 1}_1}},$ where $\delta^{1\leftarrow z_1}$ is the average number of incoming stubs from $z_1$ to infected nodes. We can proceed similarly for $X_{N}^{\boldsymbol{j}}$, $X_{N}^{\boldsymbol{k}}$, $X_{N}^{\boldsymbol{l}}$, $X_{N}^{\boldsymbol{m}}$, and
$X_{N}^{\boldsymbol{o}}$. This step requires calculating the limits for $\Theta_{(z_1,0)}$, $\Omega^{(N)}_{(z_1,0)}$, $\Gamma^{(N)}_{(z_1,0)}$, $\Gamma^{(N)}_{(z_1,1)}$ which are all equal to zero. Then, using  Slutsky’s Theorem, we have that:
\begin{flalign}
X_N \xrightarrow{D}&
\frac{\prod_{r=1}^{d^{z_1\leftarrow 1}_1}D_{i_r}^{1\rightarrow z_1}\mathds{1}_{[z_{i_r}=1]}}{(\delta^{1\leftarrow z_1})^{d^{z_1\leftarrow 1}_1}}
\cdot \frac{\prod_{r=1}^{d^{z_1\leftarrow 0}_1}D_{j_r}^{0\rightarrow z_1}\mathds{1}_{[z_{j_r}=0]}}{(\delta^{0\leftarrow z_1})^{d^{z_1\leftarrow 0}_1}}\cdot\notag\\
&\frac{\prod_{r=1}^{d^{z_1\rightarrow 1}_1}D_{k_r}^{1\leftarrow z_1}\mathds{1}_{[z_{k_r}=1]}}{(\delta^{1\rightarrow z_1})^{d^{z_1\rightarrow 1}_1}}
\cdot \frac{\prod_{r=1}^{d^{z_1\rightarrow 0}_1}D_{l_r}^{0\leftarrow z_1}\mathds{1}_{[z_{l_r}=0]}}{(\delta^{0\rightarrow z_1})^{d^{z_1\rightarrow 0}_1}}\cdot\notag\\
&\frac{\prod_{r=1}^{d^{z_1\leftrightarrow 1}_1}D_{m_r}^{1\leftrightarrow z_1}\mathds{1}_{[z_{m_r}=1]}}{(\delta^{1\leftrightarrow z_1})^{d^{z_1\leftrightarrow 1}_1}}
\cdot \frac{\prod_{r=1}^{d^{z_1\leftrightarrow 0}_1}D_{n_r}^{0\leftrightarrow z_1}\mathds{1}_{[z_{n_r}=0]}}{(\delta^{0\leftrightarrow z_1})^{d^{z_\leftrightarrow 0}_1}}.
\end{flalign}
Therefore, the second condition of Lemma 3 is satisfied. Finally, since $D^{z_1\leftarrow 1}_{\boldsymbol{i}}$, $D^{z_1\leftarrow 0}_{\boldsymbol{j}}$, $D^{z_1\rightarrow 1}_{\boldsymbol{k}}$, $D^{z_1\rightarrow 0}_{\boldsymbol{l}}$, $D^{z_1\leftrightarrow 1}_{\boldsymbol{m}}$, $D^{z_1\leftrightarrow 0}_{\boldsymbol{o}}$ are independent by construction, we also have that the third condition of Lemma 3 is satisfied, that is:
\begin{flalign}
E&\left(
\frac{
\prod_{r=1}^{d^{z_1\leftarrow 1}_1}
D_{i_r}^{1\rightarrow z_1}
\mathds{1}_{[z_{i_r}=1]}
}
{
(\delta^{1\rightarrow z_1})^{d^{z_1\leftarrow 1}_1}
} \cdot
\frac{
\prod_{r=1}^{d^{z_1\leftarrow 0}_1}
D_{j_r}^{0\rightarrow z_1}
\mathds{1}_{[z_{j_r}=0]}
}
{
(\delta^{0\rightarrow z_1})^{d^{z_1\leftarrow 0}_1}
}\cdot\right.\notag\\
&\frac{
\prod_{r=1}^{d^{z_1\rightarrow 1}_1}
D_{k_r}^{1\leftarrow z_1}
\mathds{1}_{[z_{k_r}=1]}
}
{
(\delta^{1\leftarrow z_1})^{d^{z_1\rightarrow 1}_1}}\cdot 
\frac{
\prod_{r=1}^{d^{z_\rightarrow 0}_1}
D_{l_r}^{0\leftarrow z_1}
\mathds{1}_{[z_{l_r}=0]}
}
{
(\delta^{0\leftarrow z_1})^{d^{z_\rightarrow 0}_1}
}\cdot \notag\\
&\left.
\frac{
\prod_{r=1}^{d^{z_1\leftrightarrow 1}_1}
D_{m_r}^{1\leftrightarrow z_1}
\mathds{1}_{[z_{m_r}=1]}
}
{
(\delta^{1\leftrightarrow z_1})^{d^{z_1\leftrightarrow 1}_1}
} \cdot 
\frac{
\prod_{r=1}^{d^{z_1\leftrightarrow 0}}D_{n_r}^{0\leftrightarrow z_1}\mathds{1}_{[z_{n_r}=0]}
}
{
(\delta^{0\leftrightarrow z_1})^{d^{z_1\leftrightarrow 0}_1}
}
\right)=1.
\end{flalign}
Therefore, we use Lemma 3 to conclude that:
\begin{flalign}
\lim_{N\to \infty}
E&\left(
\frac{
\prod_{r=1}^{d^{z_1\leftarrow 1}_1}
D_{i_r}^{1\rightarrow z_1}\mathds{1}_{[z_{i_r}=1]}
}
{
(\delta^{1\rightarrow z_1})^{d^{z_1\leftarrow 1}_1}
}
\frac{
\prod_{r=1}^{d^{z_1\leftarrow 0}_1}
D_{j_r}^{0\rightarrow z_1}
\mathds{1}_{[z_{j_r}=0]}
}
{
(\delta^{0\rightarrow z_1})^{d^{z_1\leftarrow 0}_1}
}\cdot 
\right.\notag\\
&\frac{
\prod_{r=1}^{d^{z_1\rightarrow 1}_1}
D_{k_r}^{1\leftarrow z_1}
\mathds{1}_{[z_{k_r}=1]}
}
{
(\delta^{1\leftarrow z_1})^{d^{z_1\rightarrow 1}_1}
}
\frac{
\prod_{r=1}^{d^{z_1\rightarrow 0}_1}
D_{l_r}^{0\leftarrow z_1}
\mathds{1}_{[z_{l_r}=0]}
}
{
(\delta^{0\leftarrow z_1})^{d^{z_1\rightarrow 0}_1}
}\notag\\
&\left.
\frac{
\prod_{r=1}^{d^{z_1\leftrightarrow 1}_1}
D_{m_r}^{1\leftrightarrow z_1}
\mathds{1}_{[z_{m_r}=1]}
}
{
(\delta^{1\leftrightarrow z_1})^{d^{z_1\leftrightarrow 1}_1}
}
\frac{
\prod_{r=1}^{d^{z_1\leftrightarrow 0}_1}
D_{n_r}^{0\leftrightarrow z_1}
\mathds{1}_{[z_{n_r}=0]}
}
{
(\delta^{0\leftrightarrow z_1})^{d^{z_1\leftrightarrow 0}_1}
}
\right)=1,
\end{flalign}
which implies that:
\begin{align}
    \lim_{N\to \infty}& P(M^{(N^{z_1})}_{1}=0|\boldsymbol{D}_1=\boldsymbol{d}_1)= \lim_{N\to \infty} c\cdot E\left[X_N \right]= 1. \hspace{3cm} \qedsymbol{} \notag
\end{align}

\section{Algorithm to generate the network} \label{network_generation}
This section describes the algorithm used in the simulation study to generate networks. In this algorithm, the adjacency matrix $Y$ is partitioned based on the infection status of the nodes. Therefore, $Y$ is divided into four blocks: relationships among infected nodes (block $Y_{11}$), uninfected nodes (block $Y_{00}$), and between infected and uninfected nodes (blocks $Y_{10}$ and $Y_{01}$). The probability of a connection between two nodes varies by block and is determined by the population size $N$ and the parameters $w$, $m$, $h$, $\phi$, and $\lambda$. Also, the proportion of directed edges is controlled by the parameter $\alpha$. 

The algorithm is as follows:
\begin{enumerate}
\item[(a)] Determine the number of edges in each of the four blocks of the simulated adjacency matrix ($E_{11}^s$, $E_{01}^s$, $E_{10}^s$, $E_{00}^s$) based on Equations (31) to (34) of the main article and on $N$, $w$, $m$, $h$, $\phi$, and $\lambda$.
\item[(b)]  Determine the total number of edges ($TE^s$) in $Y$ such that:
\begin{align}
TE^s=E_{11}^s+E_{10}^s+E_{01}^s+E_{00}^s
\end{align}

\item[(c)]  Calculate the number of directed ($DE^s$) and undirected ($UE^s$) edges using the parameter $\alpha$, Equations (\ref{eq:DE}) and (\ref{eq:UE}), and round the results to the nearest integer:
\begin{align}
DE^s &= \alpha \cdot TE^s\label{eq:DE}\\
UE^s &=(1-\alpha) \cdot TE^s. \label{eq:UE}
\end{align}

\item[(d)] Check if $E_{11}^s+E_{00}^s+2 \cdot min\{E_{10}^s+E_{01}^s\} > UE^s$. If not, select a higher value of $\alpha$ and return to step 3.

\item[(e)]  Calculate the total number of possible undirected edges ($PUE^s$) as follows:
\begin{align}
PUE^s=E_{11}^s+E_{00}^s+2\cdot \min\left\{E_{10}^s, E_{01}^s\right\},
\end{align}
and the number of undirected edges by block as follows:
\begin{align}
UE_{11}^s&=\frac{UE^s}{PUE^s} \cdot E_{11}^s  \hspace{3.2cm} (\text{block } Y_{11}) \notag\\
UE_{00}^s&=\frac{UE^s}{PUE^s} \cdot E_{00}^s  \hspace{3.2cm} (\text{block } Y_{00}) \notag\\
UE_{10}^s&=\frac{UE^s}{PUE^s}\cdot  2\cdot \min \{E_{10}^s,E_{01}^s\} \hspace{1cm} (\text{blocks } Y_{10} \text{ and } Y_{01}).
\notag
\end{align}

\item[(f)] For blocks $Y_{kk}$ where $k = 0, 1$, simulate edges at random with probabilities equal to $p_{kk}$. To ensure that the total number of undirected edges equals $UE_{kk}^s$ by randomly reordering the edges within the blocks until $UE_{kk}^s$ is achieved.

\item[(g)] For blocks $Y_{10}$ and $Y_{01}$, simulate edges at random with probabilities equal to $p_{10}$, and $p_{01}$, respectively. Make sure that the total number of undirected edges equals $UE_{10}^s$ by randomly reordering the edges within one of the two blocks until $UE_{10}^s$ is achieved.

\end{enumerate}

\section{Additional results} \label{more_results}

In this section, we present additional results from our simulation study presented in the main article.

\begin{figure}[!ht]%
\centering
\includegraphics[width = 11cm]{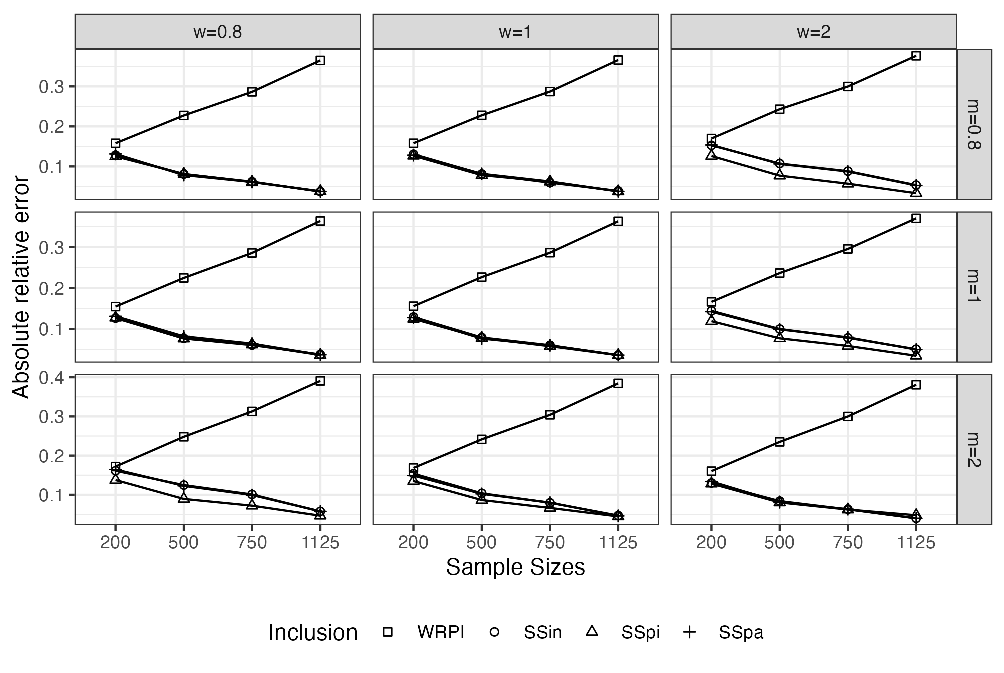}
\caption{Mean absolute relative error of four approximations of the inclusion probabilities for RDS ($WRPI$, $SS_{in}$, $SS_{pi}$, $SS_{pa}$)  for nine network scenarios ($w, m \in \{0.8, 1, 2\}$) with sample sizes ($n\in \{200,500,750,1125\}$), with low homophily ($h=1$) and low proportion of directed edges ($\alpha=0.2$).}
\label{figura_MSE1_h1}
\end{figure}

Figure~\ref{figura_MSE1_h1} compares networks without homophily ($h=1$) using the mean absolute relative error (MARE) defined by:
\begin{equation}\label{MARE}
    MARE=\frac{1}{N}\sum_{i=1}^N\frac{|\hat{\pi}^{app}_i-\hat{\pi}^{rds}_{i}|}{\hat{\pi}^{rds}_i}.
\end{equation} 
where $\hat{\pi}_i^{app}$ is the relative frequency of each node across the samples of a given scenario. The similarity with the $RDS$ sampling probabilities increases for larger sample sizes for all approximations except WRPI. Again, $SS_{pi}$ has the best performance across all scenarios and sample sizes, closely followed by  $SS_{pa}$ and $SS_{in}$. 

\begin{figure}[!ht]%
\centering
\includegraphics[width = 11cm]{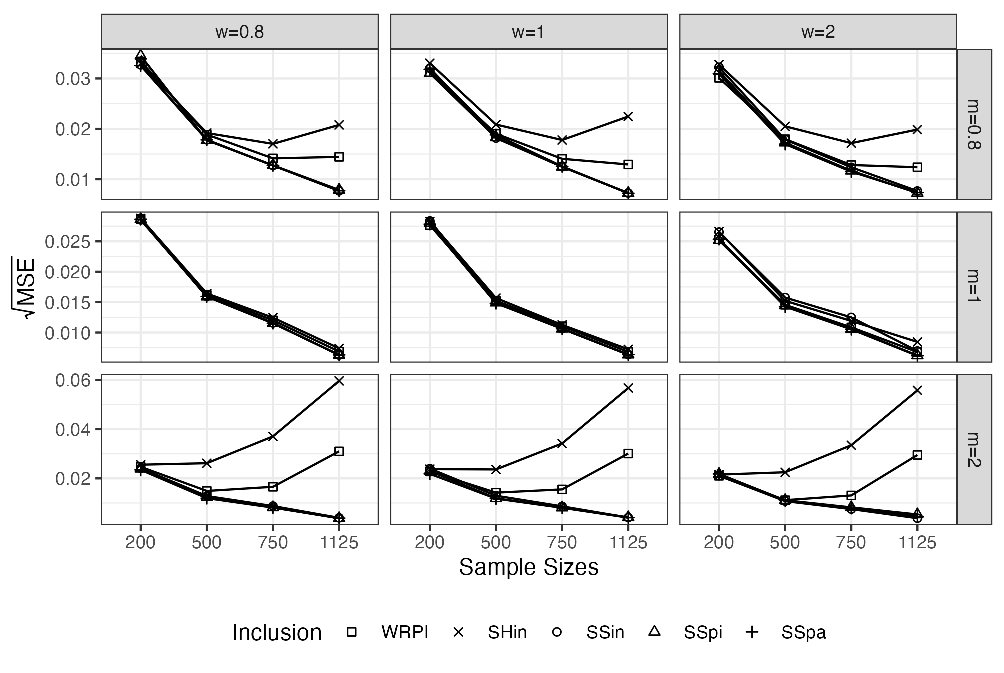}
\caption{The root mean square error for the prevalence estimators with different inclusion probabilities given by the approximation $WRPI$, $SS_{in}$, $SS_{pi}$ and $SS_{pa}$. The figure also includes the $SH_{in}$ estimator. The proportion of directed edges is 0.2, and the homophily is equal to 1.}
\label{MSE_h1}
\end{figure}

Figure \ref{MSE_h1} shows the Root Mean square error (RMSE) for the prevalence estimation when the homophily equals 1. The RMSE for the estimators using our proposed RDS approximations and the $SS_{in}$ decreases when the sample size increases. However, this trend is not always seen for the estimators based on the WRPI representation since they are affected by the finite population effect.

\begin{figure}[!ht]%
\centering
\includegraphics[width = 11cm]{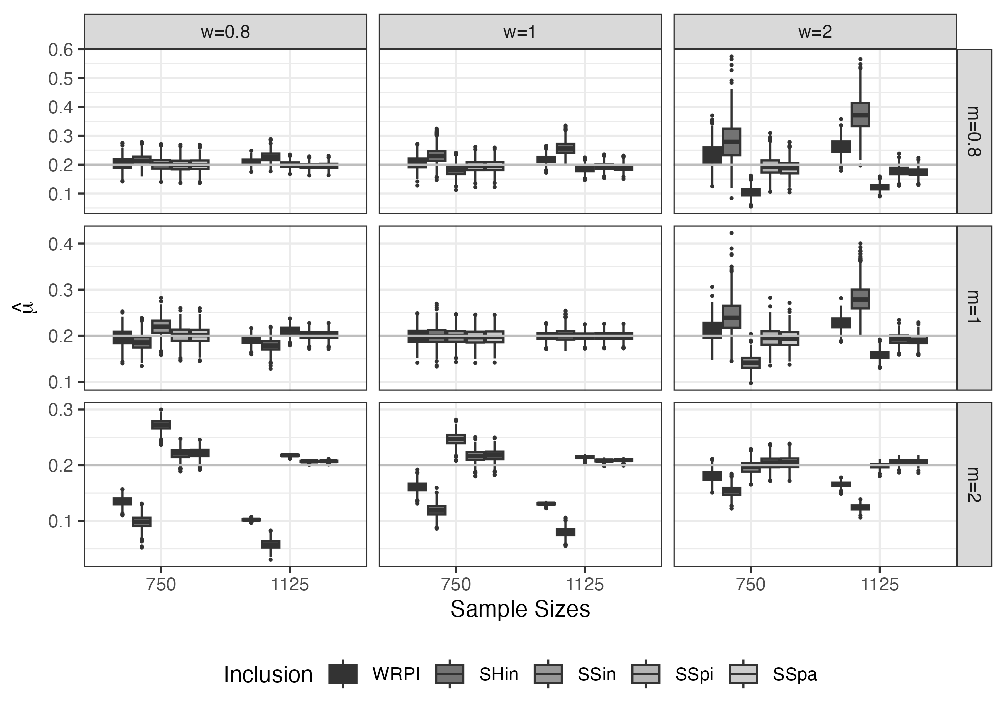}
\caption{Hájek's estimator based on WRPI, $SS_{in}$, $SS_{pi}$, $SS_{pa}$ and the $SH_{in}$ estimator when the network has different values of attractiveness $m$  and activity $w$ ratios, for samples of sizes 750 and 1125. The homophily is high ($h=5$), and the proportion of directed edges is 0.2.}
\label{boxplot1.2}
\end{figure}

Figure \ref{boxplot1.2} summarizes the approximated sampling distribution of the prevalence estimators for RDS samples of sizes 750 and 1125 and homophily equal to five. When the sample size increases, the bias of all SS-based estimators decreases while it increases for the WRPI-based estimators due to the finite population effect. Also, our proposals perform better than the other estimators when $m\neq w$ as shown by their lower bias and variability.

\begin{figure}[!ht]%
\centering
\includegraphics[width = 11cm]{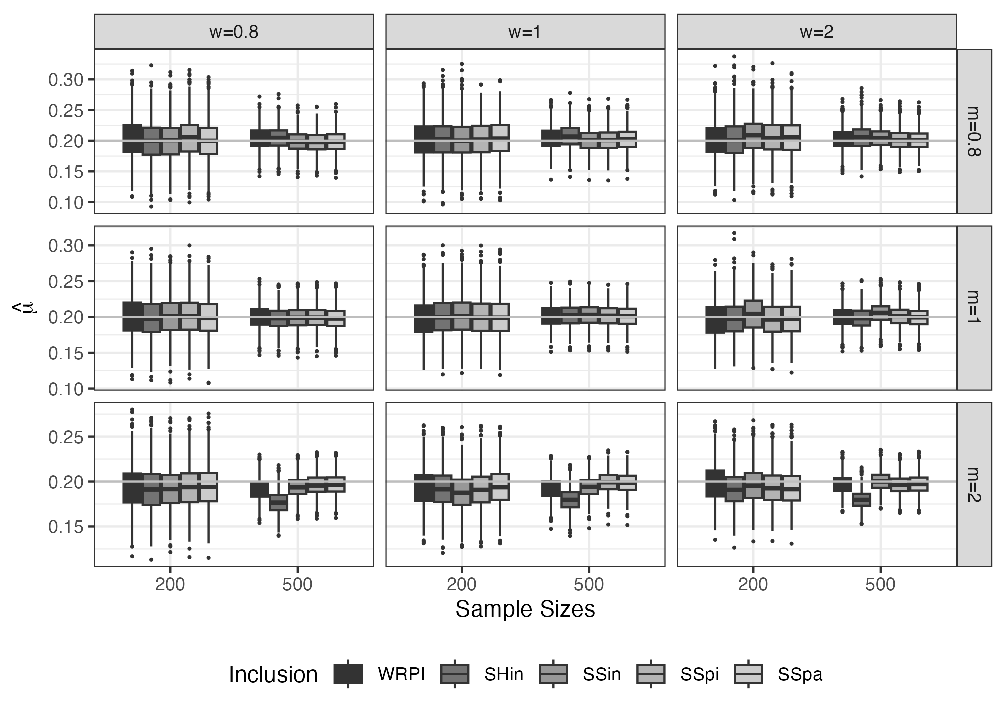}
\caption{Hájek's estimator based on $WRPI$, $SS_{in}$, $SS_{pi}$ and $SS_{pa}$ when the network have different values of attractiveness $m$  and activity $w$ ratios, for samples sizes of 200 and 500. Homophily is low ($h=1$), and the proportion of directed edges is 0.2. Also, the figure includes the $SH_{in}$ estimator.}
\label{boxplot2_h1}
\end{figure}

\begin{figure}[!ht]%
\centering
\includegraphics[width = 11cm]{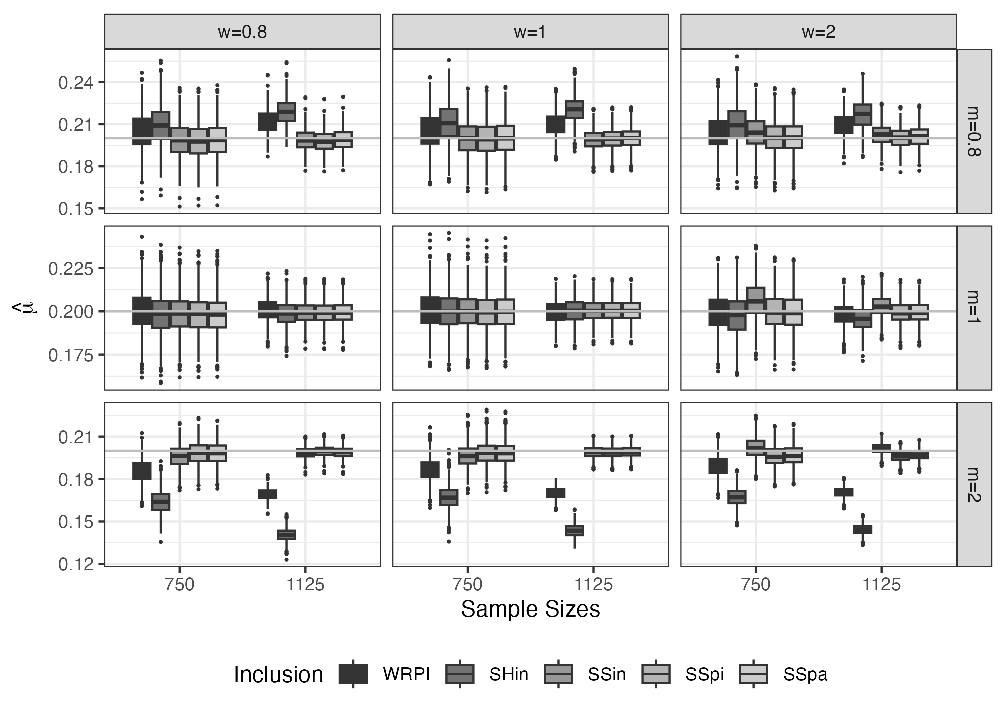}
\caption{Hájek's estimator based on $WRPI$, $SS_{in}$, $SS_{pi}$ and $SS_{pa}$ when the network have different values of attractiveness $m$  and activity $w$ ratios, for samples sizes of 750 and 1125. Homophily is low ($h=1$), and the proportion of directed edges is 0.2. Also, the figure includes the $SH_{in}$ estimator.}
\label{boxplot3_h1}
\end{figure}

Figure \ref{boxplot2_h1} and Figure \ref{boxplot3_h1} also summarize the approximated sampling distribution of the prevalence estimators. However, they present the information for RDS samples of sizes 200, 500 (Figure \ref{boxplot2_h1}) and sizes 750 and 1125 (Figure \ref{boxplot3_h1}) when the homophily equals one. The differences between the RDS approximations are not as pronounced as when the homophily equals five. However, we note more significant discrepancies with WRPI-based estimators for larger sample sizes.

\newpage
\bibliographystyle{plainnat}
\bibliography{supplement}